# Toroidal helical pulses

Shuai Shi[1,#], Hongcheng Zhou[2,#], Junjie Shao[1], Pan Tang[1], Bing-Zhong Wang[1], Mu-Sheng Liang[1], Yanhe Lyu[3], Boris A. Malomed[4], Yijie Shen[3,5],Ren Wang[1*]

[1] *Institute of Applied Physics, University of Electronic Science and Technology of China, Chengdu 611731, China*

[2] *Key Laboratory of Magnetic Suspension Technology and Maglev Vehicle, Ministry of Education, School of Electric Engineering, Southwest Jiaotong University, Chengdu 610031, China*

[3] *Centre for Disruptive Photonic Technologies, School of Physical and Mathematical Sciences, Nanyang Technological University, Singapore 637371, Singapore*

[4] *Instituto de Alta Investigación, Universidad de Tarapacá, Casilla 7D, Arica, Chile*

[5] *School of Electrical and Electronic Engineering, Nanyang Technological University, Singapore 639798, Singapore*

# Shuai Shi and Hongcheng Zhou contribute equally to this work.

* Corresponding author: Ren Wang. E-mail: rwang@uestc.edu.cn

## Abstract

**Toroidal topologies and helicity are pervasive in nature and hold basic importance in scientific research. In particular, the interplay between these features gives rise to fascinating toroidal helical electromagnetic excitations. Here, we present a theoretical framework and experimental realization to introduce a family of toroidal helical pulses, exploring the intersection of the helicity and propagating toroidal modes. For this purpose, we propose a configuration combining a coaxial horn emitter and an equiangular spiral grating to directly generate such single-cycle pulses. In addition to their inherent non-transverse toroidal topology and space-time nonseparability, such pulses also possess controllable helicity. This work gives rise to a helical version of propagating toroidal electrodynamics, thereby paving the way for advanced applications, such as nontrivial light-matter interactions and data transfer.**

## Introduction

Toroidal topologies [1]–[3] and helicity [4],[5] constitute two fundamental organizing principles in modern topological photonics. Over the past decade, a broad range of structured electromagnetic fields exhibiting toroidal and helical topology has been realized in both material platforms and free space. Representative examples include photonic conchs [6], polar spirals [7], vortex and helical knots [8]–[21], Möbius strips [22]–[27], and hopfions and skyrmions [28]–[40].

A significant advance in this direction was the experimental realization of free-space toroidal electromagnetic pulses [2],[41],[42]. These broadband single-cycle or few-cycle excitations exhibit strong longitudinal components and intrinsic space–time nonseparability, enabling robust propagation and unconventional light–matter interaction channels [43]–[47]. Their toroidal energy circulation distinguishes them from conventional transverse beams and has stimulated growing interest in toroidal electrodynamics [48], where toroidal multipoles, dynamic anapoles and ultrafast structured fields are treated within a unified framework. However, the mirror-symmetric toroidal topology of such pulses limits their ability to produce light-matter interactions and constrains their potential applications.

In this work, we present a theoretical framework which introduces a family of toroidal helical pulses, whose fields form a toroidal helix structure. We further propose a configuration combining a coaxial horn emitter and an equiangular spiral grating as a setup for direct generation of such single-cycle pulses. In addition to their inherent non-transverse toroidal topology and space-time nonseparability, these pulses also possess controllable helicity. Previously unobserved hybrid electromagnetic skyrmions are identified on the basis of these pulses. Thus, this work extends the family of toroidal pulses by including the helical subclass, thereby expanding their potential applications.

**Results**

**The topology of toroidal helical pulses.**

In 1996, Hellwarth and Nouchi introduced a non-transverse propagating solution of Maxwell's equations [49], now called toroidal pulses. The electric and magnetic fields of transverse electric (TE) toroidal pulses in the spatial cylindrical coordinate system ($r$, $z$) are expressed as follows:

$$E_{\varphi} = -4\mathrm{i}f_0\sqrt{\frac{\mu_0}{\varepsilon_0}}\frac{r(q_1+q_2-2\mathrm{i}ct)}{[r^2+(q_1+\mathrm{i}\tau)(q_2-\mathrm{i}\sigma)]^3} \quad (1)$$

$$H_{\mathrm{r}} = 4\mathrm{i}f_0\frac{r(q_2-q_1-2\mathrm{i}z)}{[r^2+(q_1+\mathrm{i}\tau)(q_2-\mathrm{i}\sigma)]^3} \quad (2)$$

$$H_{\mathrm{z}} = -4f_0\frac{r^2-(q_1+\mathrm{i}\tau)(q_2-\mathrm{i}\sigma)}{[r^2+(q_1+\mathrm{i}\tau)(q_2-\mathrm{i}\sigma)]^3} \quad (3)$$

where $f_0$ is a normalization constant, $t$ is time, $\sigma = z + ct$, $\tau = z - ct$, $r^2=x^2+y^2$, $q_1$ and $q_2$ being the effective wavelength and Rayleigh range, respectively.

In free space, the TM solution can be obtained from the TE solution through the dual-symmetry transformation of Maxwell's equations, where the electric and magnetic fields are interchanged as:

$$\mathbf{E}_{\mathrm{TM}} = \sqrt{\frac{\mu_0}{\varepsilon_0}}\mathbf{H}_{\mathrm{TE}}, \mathbf{H}_{\mathrm{TM}} = -\sqrt{\frac{\varepsilon_0}{\mu_0}}\mathbf{E}_{\mathrm{TE}}, \quad (4)$$

The topologies of transverse magnetic (TM) and TE toroidal pulses are identical, differing only in that their electric and magnetic fields are interchanged. Both TE and TM toroidal pulses exhibit mirror symmetry, which makes the field distribution symmetric with respect to the plane that bisects the pulse, so that the fields on one side of the plane are a mirror image of those on the other side. By superposing TE and TM toroidal pulses, we obtain a family of toroidal helical pulses, introducing helicity to this class of space-time nonseparable toroidal solutions. The electric field of an electromagnetic toroidal helical pulse can be expressed as

$$
\begin{aligned}
\mathbf{E} &= \mathrm{e}^{\mathrm{i}\beta}\cos(\alpha)\mathbf{E}_{\mathrm{TE}} + \sin(\alpha)\mathbf{E}_{\mathrm{TM}} \\
&= \mathrm{e}^{\mathrm{i}\beta}\cos(\alpha)E_{\varphi}\hat{\boldsymbol{e}}_{\varphi} + \sin(\alpha)\left(E_r\hat{\boldsymbol{e}}_r + E_z\hat{\boldsymbol{e}}_z\right),
\end{aligned} \tag{5}
$$

where $\alpha \in [-\pi/2, \pi/2]$ is an amplitude superposition factor, and $\beta$ denotes their relative phase. For $\alpha = 0$ or $\pm\pi/2$, the electromagnetic toroidal helical pulse degenerates into a TM or TE toroidal pulse, respectively. For $\beta = 0$, the traced field lines exhibit multiple closed loops; otherwise, they wind around the torus to form a single closed loop. For more detailed derivations and discussions, please refer to Supplementary Notes 1.

The topology and generation scheme of toroidal helical pulses are displayed in Fig. 1. Both the electric and magnetic field vectors of the pulse wind continuously around the same toroid, forming toroidal helices. On the front and rear sides, the theoretical electric and magnetic fields are mutually orthogonal, each forming an angle of $\pm\alpha$ with the radius of the toroid, thereby creating a single-cycle helical topological structure, as depicted in Figs. 1(a) and (b). The electric and magnetic field vectors exhibit opposite helicity. These vector fields are obtained by means of the Runge-Kutta method for the vector field line tracing [50], with the tracing starting points on the circles defined by ($r$=7$q_1$, $t$=±1.5$q_1$/c) as detailed in Supplementary Notes 2-3. The field configurations are produced as the numerical solutions of the full system of the linear Maxwell's equations in the free space, which are distinct from that of fundamental toroidal pulses explored in [2], [41]-[49], where the electric or magnetic fields encircle the magnetic or electric fields, respectively, in the TM and TE modes.

Toroidal helical pulses are generated using a coaxial horn paired with an equiangular spiral grating. As illustrated in Figs. 1(c1) and 1(c2), each slot of the grating is oriented at an angle α relative to the local radial direction. The grating structure comprises multiple sets of equiangular spirals, described by the parametric equation $r(\varphi) = r_0\,\mathrm{e}^{(\cot(\alpha)\varphi)}$. For this study, five gratings corresponding to $\alpha$ =5π/12, $\alpha$ =π/3, $\alpha$ =π/4, $\alpha$ =π/6, and $\alpha$ =π/12 were fabricated. The equiangular spiral grating was fabricated on a 1-mm-thick F4B dielectric substrate (a relative permittivity of 2.65 and a loss tangent of 0.001) with a metallic spiral-slot pattern (see Methods section and Supplementary Note 4 for details). In the coaxial horn, the electric field is radially polarized and can be decomposed in two components, polarized at angles $\alpha$ and $\alpha$-π/2 relative to the radial direction. One component is fully transmitted through the equiangular spiral grating, while the other is entirely rejected. The transmitted component can be considered as the superposition of radial and circumferential polarized components, corresponding to the transverse components of TM and TE toroidal pulses, respectively. The longitudinal components of the pulses are generated according to the Gauss's law [2], *viz.*, $E_z(r,z) = -\int_{z_0}^{z} \frac{\partial E_r(r,z')}{\partial r} dz'$, where $z_0$ is chosen as the reference point at which the field vanishes. Consequently, the toroidal helical pulses are formed by the superposition of TM and TE toroidal pulse components. By adjusting the rotational direction of the equiangular spiral grating, the emitted pulses can be shaped as either left- or right-handed ones. Furthermore, by controlling the rotation angle of the equiangular spiral grating, the relative proportions of the TM and TE toroidal pulse components in the toroidal helical pulses can be adjusted, thereby fitting the angles between the electric and magnetic fields and the radial direction.

The handedness of the emitted toroidal helical pulse is determined by the rotation direction of the equiangular spiral grating. Specifically, the sign of the spiral parameter controls the sign of the azimuthal phase introduced by the polarization–phase conversion process, thereby setting the helicity of the generated pulse. Reversing the rotation direction of the spiral grating therefore produces a pulse with opposite handedness. These two configurations correspond to mirror-symmetric realizations of the same physical mechanism rather than independent pulse-generation processes, as all other experimental conditions remain unchanged.

**Observation of toroidal helical pulses.**

Measured spatiotemporal electric field and spectrum of an electromagnetic toroidal helical pulse with a typical parameter of $\alpha = \pi/4$ is shown in Fig. 2. The measurement methods are detailed in the Methods and Supplementary Notes 4-6. To visualize the spatiotemporal structure of the pulse, we plot isosurfaces of the measured electric-field components. Specifically, Figs. 2(a1)–(a3) show the isosurfaces of the cylindrical components of the electric field, namely $E_r$, $E_\varphi$, and $E_z$, respectively. The displayed surfaces correspond to ±0.35 of the peak electric-field amplitude of the pulse, which are chosen to highlight the main lobes of the propagating pulse. Because the visualization is based on isosurfaces, the resulting plots may appear similar to a binary distribution. However, it should be emphasized that no binarization, thresholding, or saturation processing has been applied to the measured data; the apparent sharp boundaries arise solely from the isosurface rendering used for visualization. Fields $E_\varphi$ and $E_r$ vanish at the center, the $E_z$ component has its maximum at the same place, indicating the longitudinal polarization in the central area. This coexistence of $E_\varphi$ and $E_r$ components gives rise to the helicity of the vector fields, as shown in Fig. 3. All measured $E_\varphi$, $E_r$, and $E_z$ components of toroidal helical pulses exhibit single-cycle or $1\frac{1}{2}$-cycle waveforms, similar to the behavior of theoretical toroidal pulses [2].

During the propagation, the toroidal helical pulses maintain similar spectral characteristics and spatiotemporal field distributions, as detailed in Supplementary Notes 6-8. The preservation of the spectra in the course of the propagation ensues from the space-time nonseparability of the governing equation, which is further demonstrated by its isodiffracting structure in Fig. 2(b1). The spatial intensity profiles of all frequency components scale similarly along the pulse's trajectory in each cross section orthogonal to the propagation direction. The measured trajectories exhibit spectral crossings near $z = 0$, but they no longer intersect as they continue to propagate, directly implying the resilient propagation of the toroidal helical pulses. The respective characteristics can be represented by means of the *concurrence*,

$$\mathrm{Con} = \sqrt{[1 - \mathrm{Tr}(\rho_A^2)]}/\sqrt{1 - 1/n}, \tag{6}$$

and *entanglement of formation*

$$\mathrm{Eof} = -\mathrm{Tr}[\rho_A \log_2(\rho_A)]/\log_2(n) \tag{7}$$

for $n$- dimensional states [51], where $\rho_A$ is the reduced density matrix. The values of Con and Eof in Fig. 2(b2) show fluctuations near $z = 0$ and remain above 0.85 at $z >$ 0.4 m, indicating a strong isodiffracting characteristic and space-time nonseparability. The calculation of the nonseparability is detailed in Supplementary Note 10.

Measured three-dimensional spatiotemporal electric field vector trajectories of the generated toroidal helical pulses with the parameters of $q_1$=0.02 m and $q_2$=20$q_1$ are shown in Figs. 3(a1)-(a3). In all cases of Fig. 3, the field vectors of the pulse wind continuously around a toroidal surface, forming toroidal helices. As $\alpha$ decreases, the circumferential component of the electric field grows progressively stronger, demonstrating enhanced helicity. It is also evident from Fig. 3 that the transversely polarized electric field of the electromagnetic toroidal helical pulse comprises both radial and circumferential components, which are distinct from that of toroidal pulses [2], [41]-[49].

To intuitively illustrate how the helicity of toroidal helical pulses varies with $\alpha$, Figs. 3(b1)-(b3) presents their vector electric fields in the transverse plane. For all cases, the longitudinal component of the field gradually decreases from the center outward, reverses direction after passing through zero, and then gradually increases again. Meanwhile, the transverse component maintains an angle of approximately $\alpha$ relative to the circumferential direction. The position of the maximum value of the longitudinal component coincides with the zero point of the transverse component. These features indicate that the vector field of the electromagnetic toroidal helical pulse in the transverse planes exhibits a hybrid skyrmion texture, combining Neel and Bloch skyrmions, that was recently observed in ferromagnetics [45], but had not been demonstrated in electromagnetics prior to the present work. Regardless of the value of $\alpha$, the skyrmion number ($N_{sk}$) of the field in the transverse planes is close to 1, indicating a well-formed skyrmion vector field distribution. To further validate the skyrmion texture, we calculate the direction sphere of the vector field [25] corresponding to different $\alpha$. The direction of the vector field is distributed across the entire sphere, further confirming the presence of the skyrmion texture. The calculation method for $N_{sk}$ and the direction sphere are detailed in Supplementary Note 9. To quantify the evolution of helicity with $\alpha$, we employ the skyrmionic helicity definition [53]:

$$\gamma = \theta - \varphi N_{sk}\,, \tag{8}$$

where $\theta$ and $\varphi$ are the vector directions and radial direction relative to the horizontal line for one specific vector in the region. The helicity is calculated by averaging this quantity over all vectors within the skyrmion area. As shown in Fig. 3(c), the helicity decreases linearly with $\alpha$. This demonstrates that the helicity of both the electromagnetic toroidal helical pulse and its embedded skyrmion texture can be effectively controlled by tuning the rotation angle of the equiangular-spiral grating.

## Discussion

In this paper, we have presented the theoretical framework for the consideration of the toroidal helical pulses, along with the experimental method for generating these pulses, based on the coaxial horn and equiangular spiral grating. Through the theoretical analysis, simulations, and experiments, the spatiotemporal topology, spectral characteristics, and propagation properties of the toroidal helical pulses are demonstrated, and the effectiveness of the proposed generation method is corroborated. These pulses retain their non-transverse toroidal topology and isodiffracting characteristics, while also exhibiting the helicity and hybrid skyrmion topology. Actually, the hybrid skyrmion topology is demonstrated in the present work, numerically and experimentally.

The proposed method relies upon decomposing TE or TM toroidal pulses into reflected and transmitted electromagnetic toroidal helical pulse components, using the equiangular spiral grating. This approach is universally applicable across different frequency bands. Toroidal pulses have already been generated in optical and terahertz frequency ranges [2], [41], the equiangular spiral gratings serving as effective electromagnetic control devices in these bands. Applying the method outlined in the present work, the irradiation of the equiangular spiral gratings with toroidal pulses at the optical and terahertz frequencies holds great potential for generating toroidal helical pulses in these frequency ranges. Similar topological structures may also arise in other wave systems, such as macroscopic hydrodynamic flows, although the underlying physical mechanisms differ from those in electromagnetic systems.

Mirror-symmetric toroidal pulses exhibit topologically protected disturbance-resilient propagation characteristics [42], and, as the toroidal helical pulses can be decomposed into TE and TM toroidal pulses, they are also expected to maintain topologically protected propagation properties. The skyrmion textures identified here correspond to specific spatiotemporal slices of the propagating toroidal helical pulse. Under linear Maxwell dynamics in free space, the field evolution remains continuous and the global topological structure is preserved in the absence of singular field degeneracies where the field amplitude vanishes. While this protection refers to robustness under perturbations within the linear propagation regime, the geometric details of the skyrmion texture may still evolve smoothly during propagation without altering its global topological character.

Furthermore, toroidal helical pulses may interact with natural structures or metamaterials, especially those featuring toroidal topologies [49], exhibiting topologically protected nontrivial interactions. In particular, interactions with helical cellular and molecular structures are expected [54],[55]. Moreover, skyrmions are considered promising carriers for data transmission [56]. The hybrid skyrmion structure carried by the toroidal helical pulses, representing a previously unobserved topology of propagating electromagnetic skyrmions, suggests the use of distinct spatial polarization modes for data encoding.

In addition, the toroidal helical pulse forms a continuous three-dimensional field structure along the propagation direction, which may provide additional degrees of freedom for information encoding through controllable helicity and geometric winding governed by the spiral parameter $\alpha$. It should be noted, however, that the present structure is distinct from optical Hopfions, which are characterized by nontrivial three-dimensional field-line linking associated with a Hopf invariant. In contrast, the fields reported here arise from TE–TM toroidal mode coupling and azimuthal polarization–phase engineering, and their topology is characterized by transverse skyrmion textures and helicity rather than an explicitly constructed Hopf invariant.

More generally, electromagnetic pulses with nontrivial spatial topology can be characterized by two structural aspects: the global topology of the energy distribution and the phase–polarization structure of the field. Conventional toroidal pulses exhibit closed-loop electric and magnetic field configurations with intrinsic longitudinal components, whereas helical pulses arise from an azimuthal phase gradient producing twisted wavefronts. The toroidal helical pulses studied here retain the toroidal field topology while exhibiting helicity introduced through polarization–phase coupling induced by the equiangular spiral grating, forming a hybrid vector-field structure.

Here the helicity refers to the geometric structure of the vector field rather than optical chirality in the strict electromagnetic sense. Since the TE and TM toroidal components are superposed with zero relative phase, $\mathrm{Im}(\mathbf{E}\cdot\mathbf{B}^*) = 0$, and the generated pulses therefore do not possess finite optical chirality density.

In the context of the light-matter interaction, another promising direction is the consideration of the nonlinear propagation of the toroidal helical modes – in particular, in the form of the corresponding topological (quasi-) solitons.

Finally, higher-order toroidal pulses (supertoroidal ones) may exhibit their own attractive characteristics, such as the propagation invariance, strong longitudinal polarization, multiple skyrmion textures, and ultrafast oscillations [57]–[59]. In the course of manuscript processing, some discussions regarding the superposition of toroidal pulses were identified [60-62], and these discussions demonstrate the potential applications of superimposed toroidal pulses in axion-like fields. Applying the helicity implementation method, elaborated in the present work, to the supertoroidal pulses, the creation of supertoroidal helical pulses may be envisioned, offering the potential for the realization of novel topological textures and propagation characteristics. This may lead to sophisticated wave-matter interactions and helical carriers for the data and energy transmission.

## Methods

### The structure of the electromagnetic toroidal helical pulse generator.

The generator of the toroidal helical pulses includes the coaxial horn and equiangular spiral grating. The structure of the coaxial horn is the same as in [42]. The equiangular spiral grating, which was fabricated using printed-circuit-board technology, consists of the dielectric substrate and single-sided metal coating. The overall structure is a disk with a radius of 120 mm and a thickness of 1 mm. The dielectric substrate is made of F4B material with a relative permittivity of 2.65 and a loss tangent of 0.001. The equiangular spiral grating on the layer of metal coating is composed of 40 sets of slots, each centered around the central point of the disk. Each slot-set unit contains four elements: a long slot, a medium-length one, and two identical short slots. In each unit, the angle between the medium-length slot and the long one is 4.5°, while the angles between the two short slots and the long one are 2.25° and 6.75°, respectively. For further details, please see Supplementary Note 4.

**Measurement of toroidal helical pulses.**

A planar microwave anechoic chamber was used to measure the transversely polarized spatial electromagnetic fields produced by the electromagnetic toroidal helical pulse generator. The R&S® ZNA vector network analyzer, with the frequency range of 10 MHz to 50 GHz, was connected to both the waveguide probe and the electromagnetic toroidal helical pulse generator. The measurement system allowed the probe to move freely within the measuring space, and it also supported rotation of the antenna for the measurement along the central axis. This arrangement enabled direct measurement of the transverse components $E_r$ and $E_\varphi$. The polarization direction of the waveguide probe was adjusted to align with the radial and circumferential directions, to measure the $E_r$ and $E_\varphi$ components, respectively. The longitudinally polarized component can be determined from the transversely polarized spatial electromagnetic fields, measured according to the Gauss's law [2]. Please see Supplementary Notes 4-6 for further details of the measurement.

**Acknowledgments**

This work has been supported by the National Natural Science Foundation of China (U2341207, 62171081, 61901086). Y.S. acknowledges support from Singapore Ministry of Education (MOE) AcRF Tier 1 grants (RG157/23 & RT11/23), Singapore Agency for Science Technology and Research (A*STAR) MTO Individual Research Grants (M24N7c0080), and Nanyang Assistant Professor ship Start Up grant. B.A.M. acknowledges the support from the Israel Science Foundation through grant No. 1695/22.

**Author contributions**

R.W. conceived the ideas and supervised the project, R.W., S.S., and J.S. performed the theoretical modeling and numerical simulations, R.W. and H.Z. developed the experimental methods, S.S. and P.T. conducted the experimental measurements, R.W., Y.S., H.Z., B.Z.W., M.S.L., Y.L. and B.A.M. conducted data analysis. All authors wrote the manuscript and participated the discussions.

**Competing interests**

The authors declare no conflict of interests.

**Data availability**

The data that support the findings of this study is available from the GitHub repository at https://github.com/shishuai0cpu/-Toroidal-helical-pulses-.

**Code availability**

The code that support the findings of this study is available from the GitHub repository at https://github.com/shishuai0cpu/-Toroidal-helical-pulses-.

## Figures

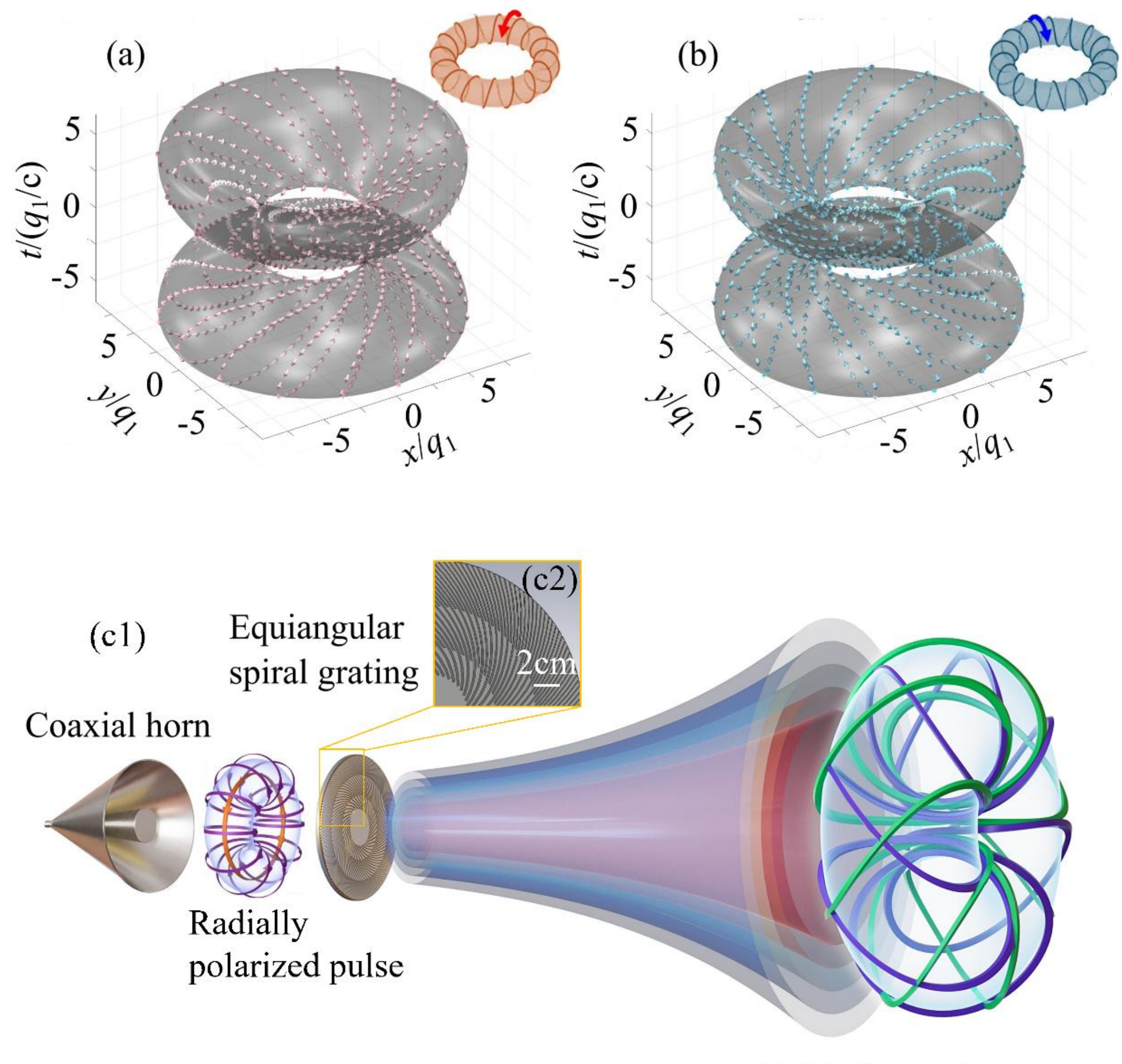

**Fig. 1 The topology and generation scheme of toroidal helical pulses.** (a) and (b): The theoretical three-dimensional vector electric and magnetic fields of an electromagnetic toroidal helical pulse with $\alpha = \pi/4$, respectively, where $q_2 = 20q_1$ and $z = 0$. The vector fields are produced by the application of the Runge-Kutta method for vector field line tracing, with the tracing starting points belonging to the circles defined by ($r=7q_1$, $t=\pm 1.5q_1/c$). Both the electric and magnetic field vectors of the pulse wind continuously around the same toroid, forming toroidal helices, similar to the coil shown in the upper-right inset. Both fields exhibit helicity, but with opposite rotation directions. The gray translucent structure in the figures represents the closed surface formed by the field tracing lines. The closed surface structures formed by the electric and magnetic

field vectors are identical. The toroidal helical pulses can be generated by using the coaxial horn and equiangular spiral grating, where each slot forms an angle of α with the radii, as shown in Fig. 1(c1). The coaxial horn emits a radially polarized toroidal electromagnetic pulse, schematically illustrated in Fig. 1(c1), where the flying-ring pattern indicates the radial polarization distribution of the incident field. A portion of the fabricated equiangular spiral grating is shown in (c2). One of the $\alpha$ and $(\alpha-\pi/2)$ polarized components is fully transmitted through the equiangular spiral grating, while the other is entirely rejected. The transmitted component can be considered as the superposition of radial and circumferential polarized components, corresponding to the transverse components of TM and TE toroidal pulses, respectively. By changing the rotation direction of the equiangular spiral grating, the emitted pulses can be adjusted to be left- or right-handed ones.

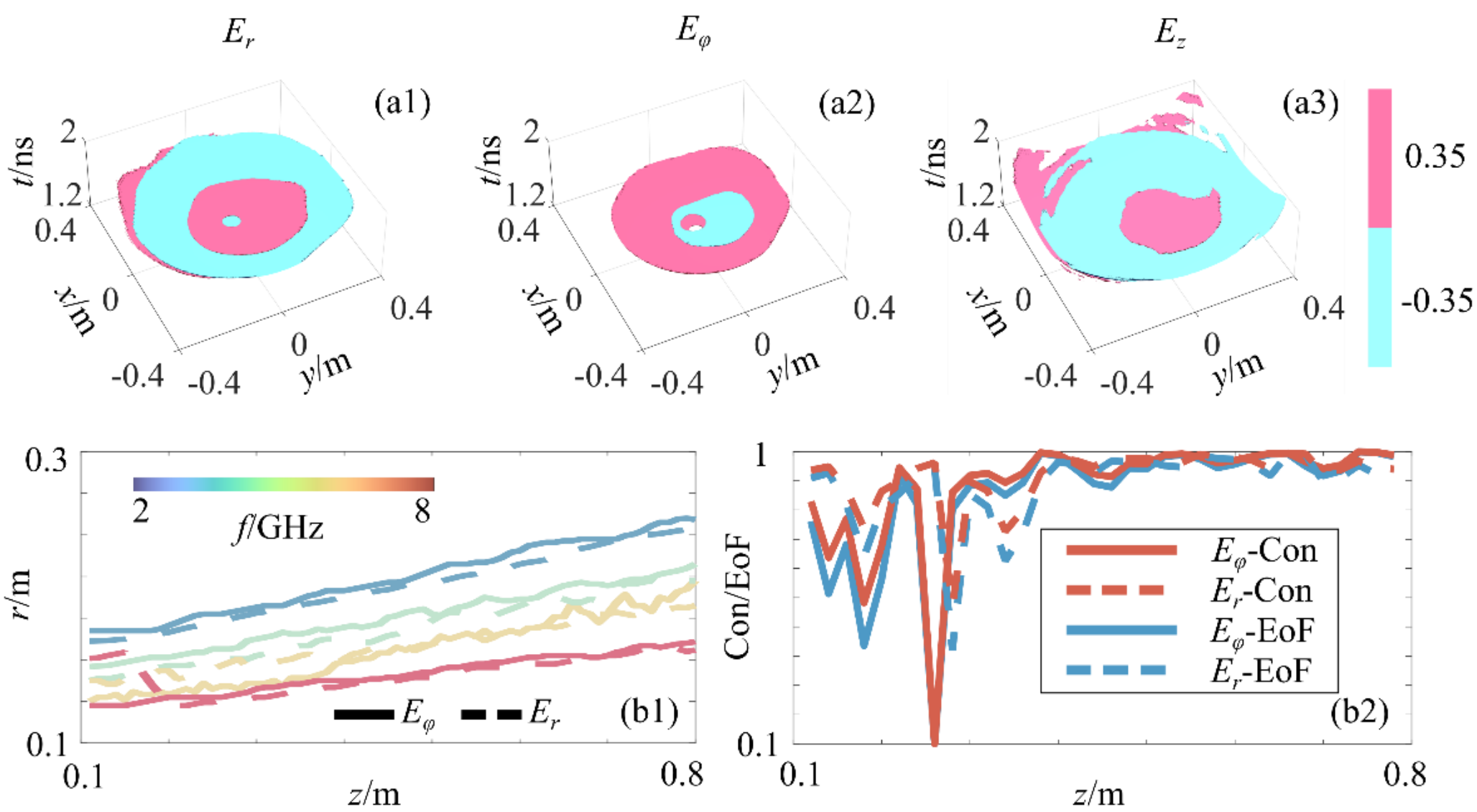

**Fig. 2 Measured spatiotemporal electric field and space-time nonseparability.** The corresponding parameters are $q_1$=0.02 m, $q_2$=20$q_1$, and $\alpha$ = π/4 and the results are plotted in the plane of $z$ = 40cm. (a1)–(a3) Isosurfaces of the cylindrical electric-field components $E_r$, $E_\varphi$, and $E_z$, respectively. The isosurfaces correspond to ±0.35 of the peak electric-field amplitude of the pulse, highlighting the main lobes of the propagating structure. Fields $E_\varphi$ and $E_r$ vanish at the center, the $E_z$ component has its maximum at the same place, indicating the longitudinal polarization in the central area.

This coexistence of $E_\varphi$ and $E_r$ components gives rise to the helicity of the vector fields. All measured $E_\varphi$, $E_r$, and $E_z$ components of toroidal helical pulses exhibit single-cycle or $1\frac{1}{2}$-cycle waveforms. The trajectories of the maximum spectrum of different components of the generated toroidal pulse during the propagation are shown in (b1). The trajectories of the $E_\varphi$ and $E_r$ components exhibit similar behavior and remain non-intersecting (except for the results near $z$ = 0), demonstrating isodiffracting characteristics, which are directly evaluated in panel (b2) in terms of the concurrence (Con) and entanglement of formation (Eof), see Eqs. (5) and (6). At $z$ > 0.4 m, the Con and Eof values of the generated toroidal helical pulses remain above 0.85.

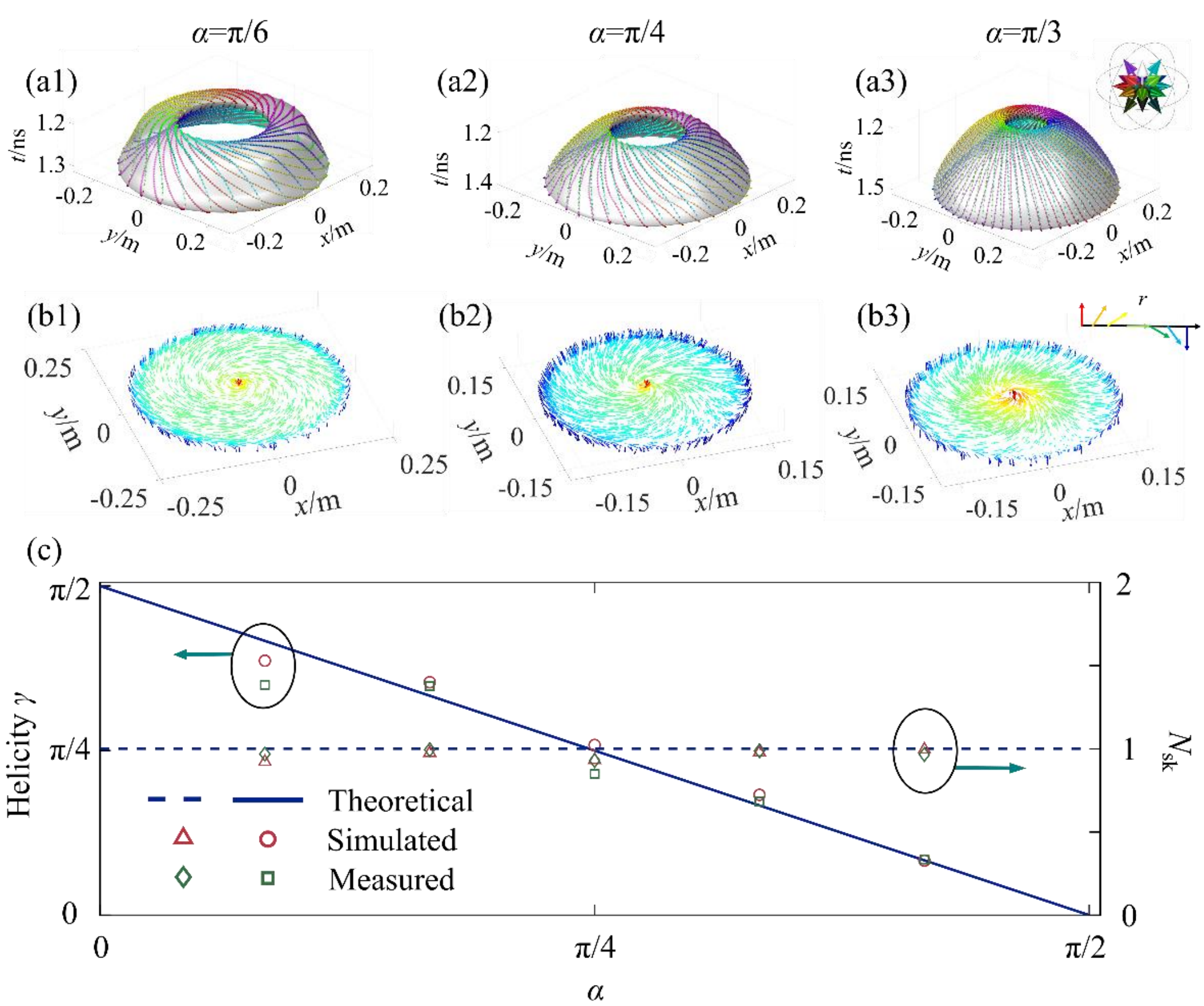

**Fig. 3 Measured spatiotemporal electric fields of toroidal helical pulses.** The parameters of toroidal helical pulses $q_1$=0.02 m and $q_2$=20$q_1$. (a1), (a2), and (a3) correspond to $\alpha$ =π/6, $\alpha$ =π/4, and $\alpha$ =π/3, respectively. In all cases, the field vectors of the pulse wind continuously around a toroidal surface, forming toroidal helices. As $\alpha$

decreases, the circumferential component of the electric field grows progressively stronger, demonstrating enhanced helicity. The transverse-plane electric vector fields corresponding to (a1), (a2), and (a3) are shown in (b1), (b2), and (b3), respectively. The vector fields are shown in the plane of $z$ = 40 cm at $t$ = 1.38 ns. The vector field of the electromagnetic toroidal helical pulse in the transverse planes exhibits a hybrid skyrmion texture, combining Neel and Bloch skyrmions. As shown in (c), regardless of the value of $\alpha$, the skyrmion number ($N_{sk}$) of the field in the transverse planes is close to 1, indicating a well-formed skyrmion vector field distribution; the helicity decreases linearly with $\alpha$. This demonstrates that the helicity of both the electromagnetic toroidal helical pulse and its embedded skyrmion texture can be effectively controlled by tuning the rotation angle of the equiangular-spiral grating.

# Supplementary Materials for

## Toroidal helical pulses

Shuai Shi[1,#], Hongcheng Zhou[2,#], Junjie Shao[1], Pan Tang[1], Bing-Zhong Wang[1], Mu-Sheng Liang[1], Yanhe Lyu[3], Boris A. Malomed[4], Yijie Shen[3,5],Ren Wang[1*]

[1] *Institute of Applied Physics, University of Electronic Science and Technology of China, Chengdu 611731, China*

[2] *Key Laboratory of Magnetic Suspension Technology and Maglev Vehicle, Ministry of Education, School of Electric Engineering, Southwest Jiaotong University, Chengdu 610031, China*

[3] *Centre for Disruptive Photonic Technologies, School of Physical and Mathematical Sciences, Nanyang Technological University, Singapore 637371, Singapore*

[4] *Instituto de Alta Investigación, Universidad de Tarapacá, Casilla 7D, Arica, Chile*

[5] *School of Electrical and Electronic Engineering, Nanyang Technological University, Singapore 639798, Singapore*

# Shuai Shi and Hongcheng Zhou contribute equally to this work.

* Corresponding author: Ren Wang. E-mail: rwang@uestc.edu.cn

**This PDF file includes:**

Supplementary Text

Figs. S1 to S15

## Supplementary Note 1: Formula derivation

The electric and magnetic fields of transverse electric (TE) toroidal pulses in the spatial cylindrical coordinate system ($r$, $z$) are expressed as follows [1]:

$$E_{\varphi} = -4\mathrm{i}f_0\sqrt{\frac{\mu_0}{\varepsilon_0}}\frac{r(q_1+q_2-2\mathrm{i}ct)}{[r^2+(q_1+\mathrm{i}\tau)(q_2-\mathrm{i}\sigma)]^3} \tag{S1}$$

$$H_r = 4\mathrm{i}f_0\frac{r(q_2-q_1-2\mathrm{i}z)}{[r^2+(q_1+\mathrm{i}\tau)(q_2-\mathrm{i}\sigma)]^3} \tag{S2}$$

$$H_z = -4f_0\frac{r^2-(q_1+\mathrm{i}\tau)(q_2-\mathrm{i}\sigma)}{[r^2+(q_1+\mathrm{i}\tau)(q_2-\mathrm{i}\sigma)]^3} \tag{S3}$$

where $f_0$ is a normalization constant, $t$ is time, $\sigma = z + ct$, $\tau = z - ct$, $r^2=x^2+y^2$, $q_1$ and $q_2$ being the effective wavelength and Rayleigh range, respectively.

In free space, the TM solution can be obtained from the TE solution through the dual-symmetry transformation of Maxwell's equations, where the electric and magnetic fields are interchanged as:

$$\mathbf{E}_{\mathrm{TM}} = \sqrt{\frac{\mu_0}{\varepsilon_0}}\mathbf{H}_{\mathrm{TE}}, \mathbf{H}_{\mathrm{TM}} = -\sqrt{\frac{\varepsilon_0}{\mu_0}}\mathbf{E}_{\mathrm{TE}}, \tag{S4}$$

Therefore, the TE and TM fields are not two unrelated constructions, but two linearly related vector realizations generated from the same complex FD solution. For this reason, the discussion below applies equally to the TE, TM, and TE+TM superposed fields.

A key result of Ref. [2] is that the real and imaginary parts form a Hilbert-transform pair. In Appendix A of that reference, the real and imaginary parts of the electric field are defined in the frequency domain as

$$\begin{aligned} E_{\mathrm{re}}(\omega) &= \frac{E(\omega)+E^*(-\omega)}{2}, \\ E_{\mathrm{im}}(\omega) &= \frac{E(\omega)-E^*(-\omega)}{2i}, \end{aligned} \tag{S5}$$

and are identified with the "1-cycle" and "1½-cycle" pulses, respectively. The same construction applies to the magnetic field. More importantly, Ref. [2] proves that

$$
\begin{aligned}
E_{\mathrm{im}}(\omega) &= i\,\mathrm{sgn}(\omega)\,E_{\mathrm{re}}(\omega), \\
H_{\mathrm{im}}(\omega) &= i\,\mathrm{sgn}(\omega)\,H_{\mathrm{re}}(\omega),
\end{aligned}
\tag{S6}
$$

with

$$
\mathrm{sgn}(\omega) = \begin{cases} 1, & \omega > 0, \\ -1, & \omega < 0, \\ 0, & \omega = 0. \end{cases}
\tag{S7}
$$

Therefore, the real and imaginary parts share the same spectrum and differ only by phase, which is precisely the defining property of a Hilbert-transform pair.

This relation immediately implies that, for all nonzero frequencies,

$$
\begin{aligned}
\left|E_{\mathrm{im}}(\omega)\right| &= \left|E_{\mathrm{re}}(\omega)\right|, \\
\left|H_{\mathrm{im}}(\omega)\right| &= \left|H_{\mathrm{re}}(\omega)\right|,
\end{aligned}
\tag{S8}
$$

Hence, the real and imaginary parts share the same spectral amplitude, bandwidth, and spatiotemporal spectral support, differing only in spectral phase: the phase shift is $+\pi/2$ for positive frequencies and $-\pi/2$ for negative frequencies, consistent with Ref. [2]. Therefore, for properties governed by spectral amplitude, bandwidth, spatiotemporal coupling, focusing, or TE/TM-coupled topology, studying either part is physically equivalent. The two are uniquely related by the Hilbert transform and should thus be viewed not as distinct pulse families, but as orthogonal representations of the same analytic electromagnetic solution. Although their instantaneous time-domain profiles and propagation evolution may differ, their underlying spectral and propagation content is identical.

More generally, the TE+TM mode can be written as

$$
\begin{aligned}
\mathbf{E} &= \mathrm{e}^{\mathrm{i}\beta}\cos(\alpha)\mathbf{E}_{\mathrm{TE}} + \sin(\alpha)\mathbf{E}_{\mathrm{TM}} \\
&= \mathrm{e}^{\mathrm{i}\beta}\cos(\alpha)E_{\varphi}\hat{\boldsymbol{e}}_{\varphi} + \sin(\alpha)\left(E_{r}\hat{\boldsymbol{e}}_{r} + E_{z}\hat{\boldsymbol{e}}_{z}\right)
\end{aligned}
\tag{S9}
$$

where, $\alpha\in[-\pi/2,\pi/2]$ represents the relative amplitude weighting of the two modes, and $\beta$ denotes their relative phase. Since Maxwell's equations satisfy the superposition principle in linear media, the field obtained after amplitude weighting and phase modulation is still a valid solution.

From a time-domain perspective, $\beta$ characterizes the relative temporal phase: when $\beta$=0, the two modes add in phase; when $\beta=\pm\pi/2$, they differ by a quarter period; other values correspond to a general phase delay.

From the perspective of polarization characterization, $\beta$ determines the phase difference between the two orthogonal components, while $\alpha$ determines their relative amplitudes; together, they define the polarization state of the resultant field. When $\beta$=0 or π, the field is linearly polarized; when $\beta=\pm\pi/2$ and the two components have equal amplitudes (e.g., $\alpha=\pm\pi/4$), the field is circularly polarized; in all other cases, it is generally elliptically polarized. Fig. S1 presents the field distributions of $\mathrm{Re}(e^{i\beta}\mathbf{E})$ under different phase conditions, illustrating the influence of relative phase variations on the field structure. Fig. S2 presents the field-line tracing results for different values of $\beta$. When $\beta = 0$, the traced field lines form multiple closed loops, whereas when $\beta = \pi/8$, they wind around the torus to form a single closed loop.

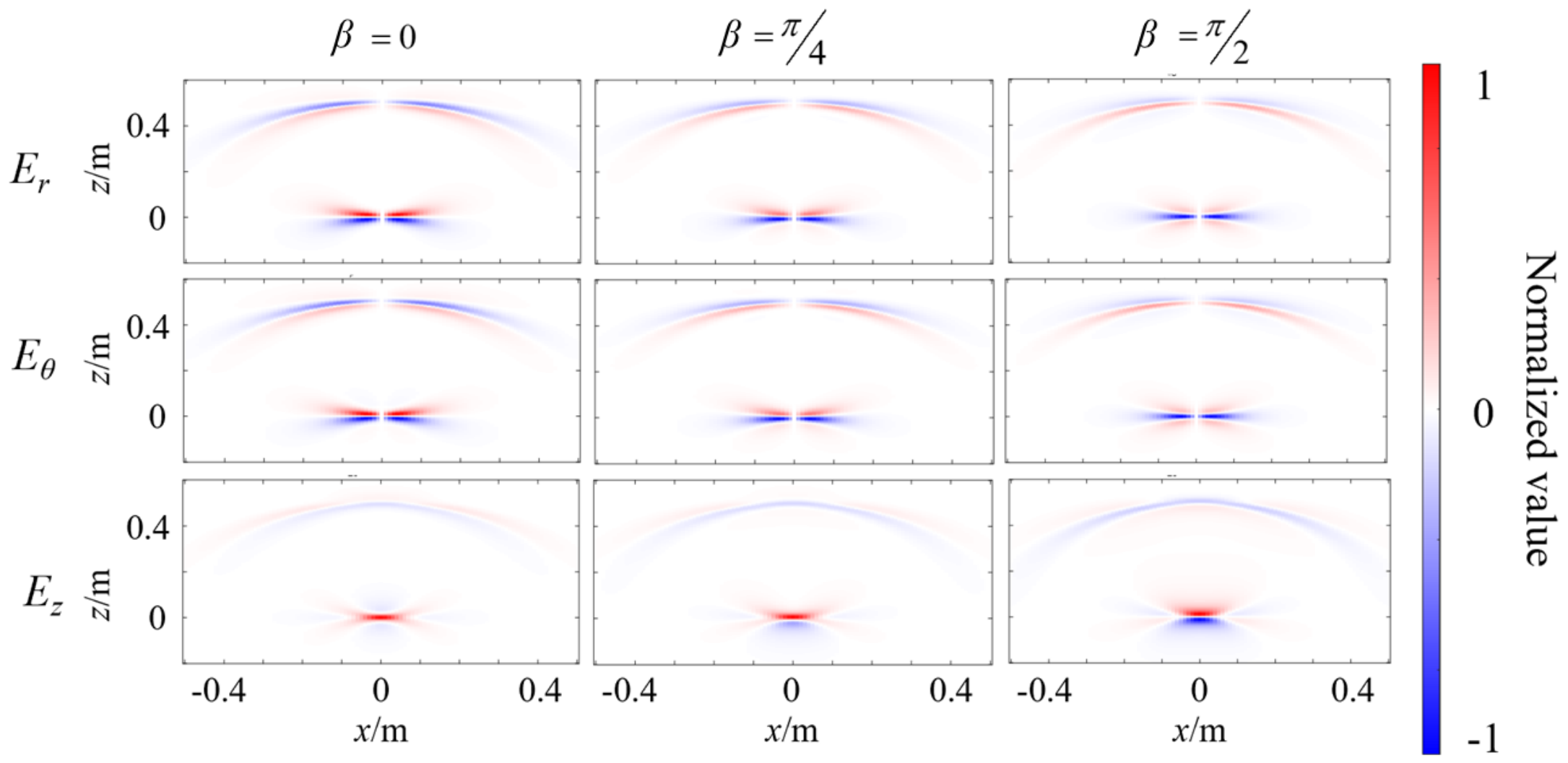

**Fig. S1 Real parts of the total field (sum of the three components) of the TE+TM flying-donut pulse for three values of $\beta$ (0, π/4 and π/2), shown at two temporal snapshots, $t$=0 and $t$=0.5/c.** The figure demonstrates that, although these fields originate from the same parent complex analytic FD solution, their instantaneous time-domain profiles depend on $\sigma$ and may exhibit different apparent cycle numbers during propagation. Under the conditions considered here, the local waveform at the observation plane can evolve into a single-cycle structure.

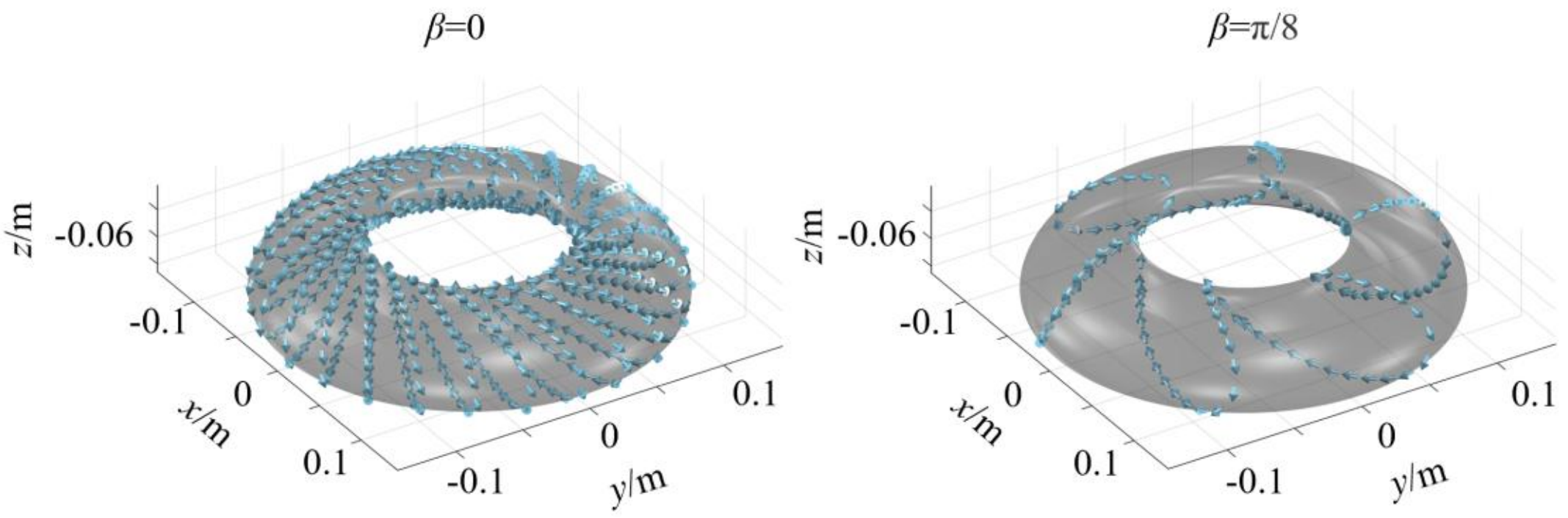

**Fig. S2 Field-line tracing results under different $\beta$ conditions.** For $\beta = 0$, the traced field lines exhibit multiple closed loops, whereas for $\beta = \pi/8$, they wind around the torus to form a single closed loop.

In this work, we mainly consider the case $\beta\approx0$, corresponding to the superposition of modes with nearly linear polarization. Although ideal linear polarization requires $\beta=0$, in practical experimental systems there usually exists a very small residual phase difference between the two orthogonal components. Therefore, choosing a small but finite $\beta$ is more consistent with experimental reality. At the same time, this treatment helps reduce the numerical sensitivity of field-line tracing under strictly degenerate conditions. Within the parameter range considered in this work, as long as $\beta$ remains sufficiently small, the resulting field distribution and its main topological features remain unchanged, and thus the physical conclusions of this paper are not affected.

**Supplementary Note 2: Methods for vector field tracking.**

Three-dimensional streamlines are a common visualization tool in fields such as fluid mechanics, representing the flow direction of a vector field at each point. These streamlines are curves whose tangents at any point align with the direction of the vector field, providing an intuitive depiction of the field's orientation across space.

Various methods exist for generating three-dimensional streamlines. In this work, the Runge-Kutta method is employed to track the electric and magnetic field vectors of the toroidal helical pulses. It is well known that this method is able to yield sufficiently accurate results [3]. Specifically, the fourth-order Runge-Kutta method is chosen for its balance between accuracy and computational efficiency, while higher-order methods can improve precision but significantly increase the computational time. The fourth-order Runge-Kutta method is defined by Eqs. (S10) and (S11):

$$
\begin{aligned}
k_1 &= V\left(P_0\right), \\
k_2 &= V\left(P_0 + \frac{\Delta s}{2}\frac{k_1}{|k_1|}\right), \\
k_3 &= V\left(P_0 + \frac{\Delta s}{2}\frac{k_2}{|k_2|}\right), \\
k_4 &= V\left(P_0 + \Delta s\frac{k_3}{|k_3|}\right),
\end{aligned}
\tag{S10}
$$

where $P_0$ represents the initial point, $\Delta s$ is the stepsize, and $V$ denotes the vector field to be traced (in this study, the electric and magnetic fields). The position of the next point, $P_1$ is determined as:

$$
P_1 = P_0 + \frac{\Delta s}{6}\left(k_1 + 2k_2 + 2k_3 + k_4\right) \tag{S11}
$$

Subsequently, $P_1$ is used as the initial point in Eqs. (S10) and (S11) to compute the next position $P_2$. This iterative process continues, with each point's position recorded, until the desired streamline length is achieved. It is important to note that the choice of stepsize $\Delta s$ significantly impacts both the accuracy and efficiency of the calculation. A stepsize that is too small increases computational load, reducing the solution speed, while an excessively large stepsize may cause streamlines to deviate from the true flow direction, compromising accuracy.

The gray translucent structures shown in Fig. S3(a) and (b) of the main text represent closed surfaces formed by the tracing lines of the electric and magnetic fields, respectively. The tracing lines are generated using the Runge-Kutta method, with starting points located on circles defined as ($r$=7$q_1$, $t$=±1.5$q_1$/c) in the vector fields of the toroidal helical pulses, as shown in Fig. S3. In Fig. S3 it is seen that the longitudinal components $E_z$ of the electric and magnetic fields of the toroidal helical pulses are identical, residing along the central longitudinal axis and the diagonal regions of the toroidal structure. The transverse components $E_r$ of the electric and magnetic fields are also identical. However, the $E_\varphi$ components have opposite directions. This relationship explains why, in Fig. 1 of the main text, both the electric- and magnetic-field vectors of the toroidal helical pulses encircle the same toroidal structure, while the electric and magnetic fields exhibit opposite rotation directions.

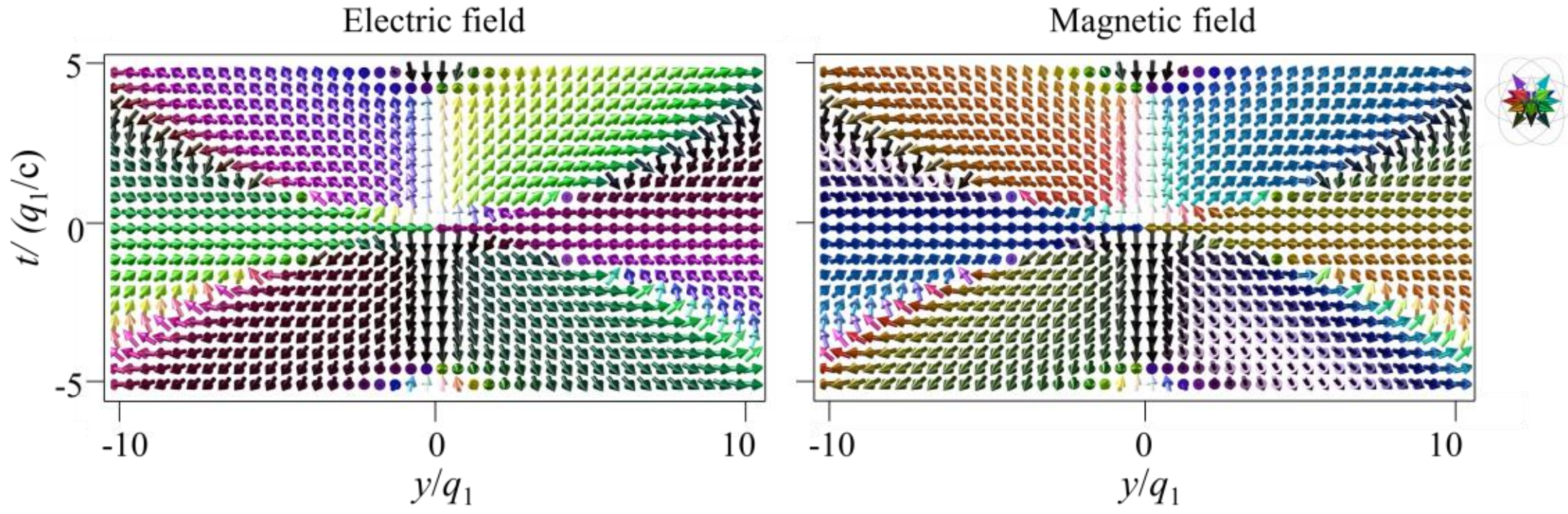

**Fig. S3 The distribution of the electric and magnetic fields of the toroidal helical pulses.** The parameters of the toroidal helical pulses are $q_2$=20$q_1$ and $z$=0. The longitudinal components $E_z$ (along the $t$ direction in this figure) of the electric and magnetic fields of the toroidal helical pulses are identical, residing along the central longitudinal axis and diagonal regions of the toroidal structure. The transverse components $E_r$ and $E_\varphi$ of the electric and magnetic fields are identical and opposite in the direction, respectively.

## Supplementary Note 3: Mapping Method

The cross-sectional curve of an irregular axisymmetric surface can be represented in

the $rz$-plane as

$$\chi(s) = \big(r(s), z(s)\big), \qquad s \in [0, L), \tag{S12}$$

where $s$ is the arc length of the cross section, and the total length is $L$.

The parameterization of the original surface is

$$S_{\text{old}}(s,\theta) = \big(r(s)\cos\theta,\, r(s)\sin\theta,\, z(s)\big), \qquad s \in [0, L), \theta \in [0, 2\pi), \tag{S13}$$

where $\theta$ is the azimuthal angle around the $z$-axis.

The parameterization of the target regular torus is

$$S_{\text{new}}(\phi,\theta) = \big((R_1 + a_1\cos\phi)\cos\theta,\, (R_1 + a_1\cos\phi)\sin\theta,\, a_1\sin\phi\big), \qquad \phi, \theta \in [0, 2\pi). \tag{S14}$$

Given an overall rotation angle $\phi_0$, define the mapping from the arc length to the circular cross-section angle as

$$\phi(s) = 2\pi\frac{s}{L} + \phi_0. \tag{S15}$$

If we want a reference position $s = s_{\text{ref}}$ on the cross section to map to a target angle $\phi_{target}$, we choose

$$\phi_0 = \phi_{\text{target}} - 2\pi\frac{s_{\text{ref}}}{L}. \tag{S16}$$

Thus a point $\mathbf{x}_{\text{old}} = S_{\text{old}}(s,\theta)$ is mapped to

$$\mathbf{x}_{\text{new}} = S_{\text{new}}(\phi(s),\theta) = \big((R_1 + a_1\cos\phi(s))\cos\theta,\, (R_1 + a_1\cos\phi(s))\sin\theta,\, a_1\sin\phi(s)\big). \tag{S17}$$

The tangent vectors of the original surface are

$$\begin{aligned} T_\theta^{\text{old}}(s,\theta) &= \frac{\partial S_{\text{old}}}{\partial\theta} = \big(-r(s)\sin\theta,\, r(s)\cos\theta,\, 0\big) \\ T_s^{\text{old}}(s,\theta) &= \frac{\partial S_{\text{old}}}{\partial s} = \big(r_s(s)\cos\theta,\, r_s(s)\sin\theta,\, z_s(s)\big) \end{aligned}, \tag{S18}$$

where $r_s = \mathrm{d}r/\mathrm{d}s$, $z_s = \mathrm{d}z/\mathrm{d}s$. The corresponding unit normal vector is defined as

$$n_{\text{old}}(s,\theta) = \frac{T_\theta^{\text{old}}(s,\theta) \times T_s^{\text{old}}(s,\theta)}{\|\, T_\theta^{\text{old}}(s,\theta) \times T_s^{\text{old}}(s,\theta) \|}. \tag{S19}$$

On the regular torus, we have

$$T_\theta^{\text{new}}(\phi,\theta)=\frac{\partial S_{\text{new}}}{\partial\theta}=\big(-(R_1+a_1\cos\phi)\sin\theta,\,(R_1+a_1\cos\phi)\cos\theta,\,0\big),$$

$$\frac{\partial S_{\text{new}}}{\partial\phi}=\big(-a_1\sin\phi\cos\theta,\,-a_1\sin\phi\sin\theta,\,a_1\cos\phi\big), \tag{S20}$$

$$\frac{\mathrm{d}\phi}{\mathrm{d}s}=\frac{2\pi}{L}.$$

Thus the tangent vector along the arc-length direction is

$$T_s^{\text{new}}(\phi,\theta)=\frac{\partial S_{\text{new}}}{\partial s}=\frac{\partial S_{\text{new}}}{\partial\phi}\frac{\mathrm{d}\phi}{\mathrm{d}s}. \tag{S21}$$

The unit normal vector on the new surface is

$$n_{\text{new}}(\phi,\theta)=\frac{T_\theta^{\text{new}}(\phi,\theta)\times T_s^{\text{new}}(\phi,\theta)}{\|\,T_\theta^{\text{new}}(\phi,\theta)\times T_s^{\text{new}}(\phi,\theta)\|}. \tag{S22}$$

Suppose at a point $\mathbf{x}_{\text{old}}=S_{\text{old}}(s,\theta)$ on the original surface there is a vector $\mathbf{E}\in\mathbb{R}^3$, In the local basis $\{T_\theta^{\text{old}}(s,\theta),T_s^{\text{old}}(s,\theta),n_{\text{old}}(s,\theta)\}$, we decompose $\mathbf{E}$ as

$$\mathbf{E}=c_\theta T_\theta^{\text{old}}(s,\theta)+c_s T_s^{\text{old}}(s,\theta)+c_n n_{\text{old}}(s,\theta). \tag{S23}$$

The normal coefficient is directly

$$c_n=\mathbf{E}\cdot n_{\text{old}}(s,\theta). \tag{S24}$$

The tangential component $\mathbf{E}_{\tan}=\mathbf{E}-\kappa n_{\text{old}}(s,\theta)$ is then expanded in the basis $\{T_\theta^{\text{old}},T_s^{\text{old}}\}$. The coefficients $c_\theta$ and $c_s$, can be obtained from the following 2×2Gram system (same as in the previously discussed tangential case):

$$\begin{bmatrix} T_\theta^{\text{old}}\cdot T_\theta^{\text{old}} & T_\theta^{\text{old}}\cdot T_s^{\text{old}} \\ [2pt]T_s^{\text{old}}\cdot T_\theta^{\text{old}} & T_s^{\text{old}}\cdot T_s^{\text{old}} \end{bmatrix}\begin{bmatrix} c_\theta \\ c_s \end{bmatrix}=\begin{bmatrix} T_\theta^{\text{old}}\cdot\mathbf{E}_{\tan} \\ [2pt]T_s^{\text{old}}\cdot\mathbf{E}_{\tan} \end{bmatrix}. \tag{S25}$$

On the target torus, at the corresponding point $\mathbf{x}_{\text{new}}=S_{\text{new}}(\phi(s),\theta)$, we reconstruct the vector using the same coefficients $(\alpha,\beta,\gamma)$ in the new basis:

$$\mathbf{E}'=c_\theta T_\theta^{\text{new}}(\phi(s),\theta)+c_s T_s^{\text{new}}(\phi(s),\theta)+c_n n_{\text{new}}(\phi(s),\theta). \tag{S26}$$

Therefore, if the original vector is purely tangential $c_n=0$, then the mapped vector $\mathbf{E}'$ remains purely tangential on the target surface. If the original vector contains a normal component, the scalar coefficient $\kappa$ remains unchanged during the mapping, while the direction follows the change of the surface normal from $n_{\text{old}}$ to $n_{\text{new}}$ , The entire process can be understood as preserving the coordinate coefficients in the local three-dimensional basis $\{T_\theta, T_s, n\}$ , thereby achieving a geometrically consistent mapping of position and tangential/normal vectors. Using the above method, we complete the mapping from an irregular toroidal surface to a regular toroidal surface. The mapping results are shown in Fig. S4

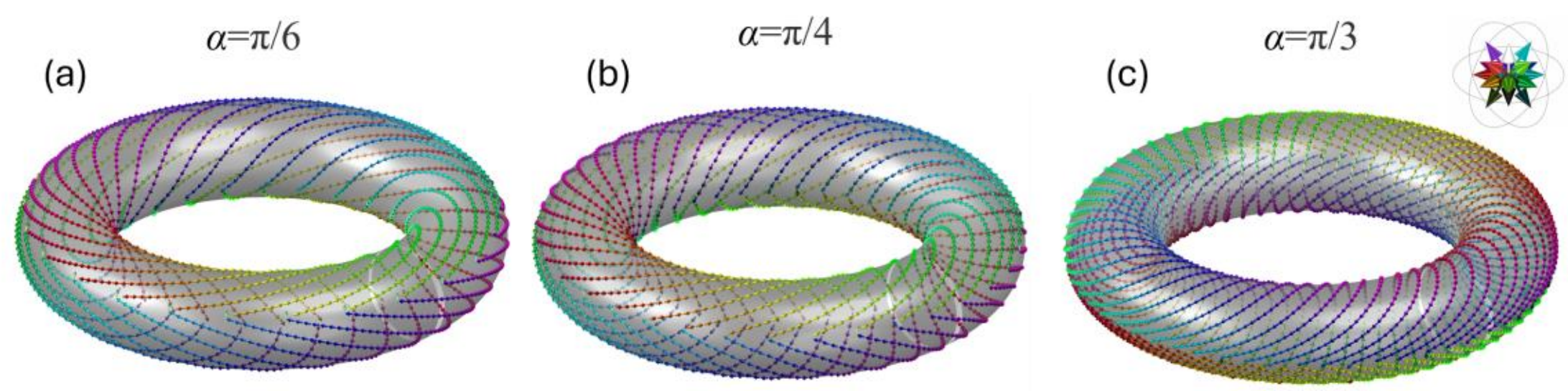

**Fig. S4 The field line mapping results at different parameter $\alpha$ are mapped from the corresponding parameter images in Figure 3.**

**Supplementary Note 4: Structures of the electromagnetic toroidal helical pulse generator.**

The generator of the toroidal helical pulses used in this work includes the coaxial horn and equiangular spiral grating. The coaxial horn consists of inner and outer conical metal conductors, fabricated by means of 3D printing, and supported by flat-shaped dielectric substrates. The horn is fed through a coaxial connector operating in a radially polarized mode. Detailed information about the coaxial horn structure can be found in the supplementary material of Ref. [4]. The following section focuses on the model and structure of the equiangular spiral grating.

The equiangular spiral grating is fabricated using the printed-circuit-board technology, consisting of a dielectric substrate and single-sided metal coating. The dielectric substrate is a circular disk with the radius of 120 mm and thickness of 1 mm, having the relative permittivity of 2.65 and loss tangent of 0.001. The metal coating on the dielectric substrate has a thickness of 0.035 mm. By etching equiangular spiral-shaped slots into the metal coating, the equiangular spiral grating used in this work is realized.

The equiangular spiral grating is composed of 40 sets of slots, each centered around the origin of the disk. Each set unit contains four slots: one long slot, one medium-length one, and two identical short slots. Two longer edges of all slots are equiangular spiral lines. Using the center of the disk as the origin of the coordinate system, the parametric equation of the equiangular spiral is given by

$$r(\varphi) = r_0 \, \mathrm{e}^{(\cot(\alpha)\varphi)} \tag{S27}$$

where $\alpha$ is the deflection angle that defines the equiangular spiral. For this study, five gratings corresponding to $\alpha = 5\pi/12$, $\alpha = \pi/3$, $\alpha = \pi/4$, $\alpha = \pi/6$, and $\alpha = \pi/12$ were fabricated, as shown in Fig. S5.

After establishing the equiangular spiral model, we set the line width to 2 mm. By trimming and removing the portion within a 40 mm radial distance from the origin, the remaining shape becomes the longest slot. Next, with the origin as the center, circles with radii of 55 mm and 90 mm are drawn and intersected with the longest slot. The uncovered portions correspond to the short slot and the medium-length slot shapes. The medium-length slot shape is then rotated 4.5 degrees about the origin, and the two short slots are rotated by 2.25 degrees and 6.75 degrees, respectively. In this way, the complete slot unit structure is obtained. Finally, using the origin as the center and 9 degrees as the rotation angle, 40 periods of spiral slots are generated and evenly distributed over the disk.

The equiangular spiral grating was then positioned closely on the mouth of the coaxial horn and fixed using an adhesive tape, ensuring that the center of the equiangular spiral grating disk remained aligned with the central axis of the coaxial horn. The installed electromagnetic toroidal helical pulse generator with the equiangular spiral grating and coaxial horn are displayed in Fig. S5.

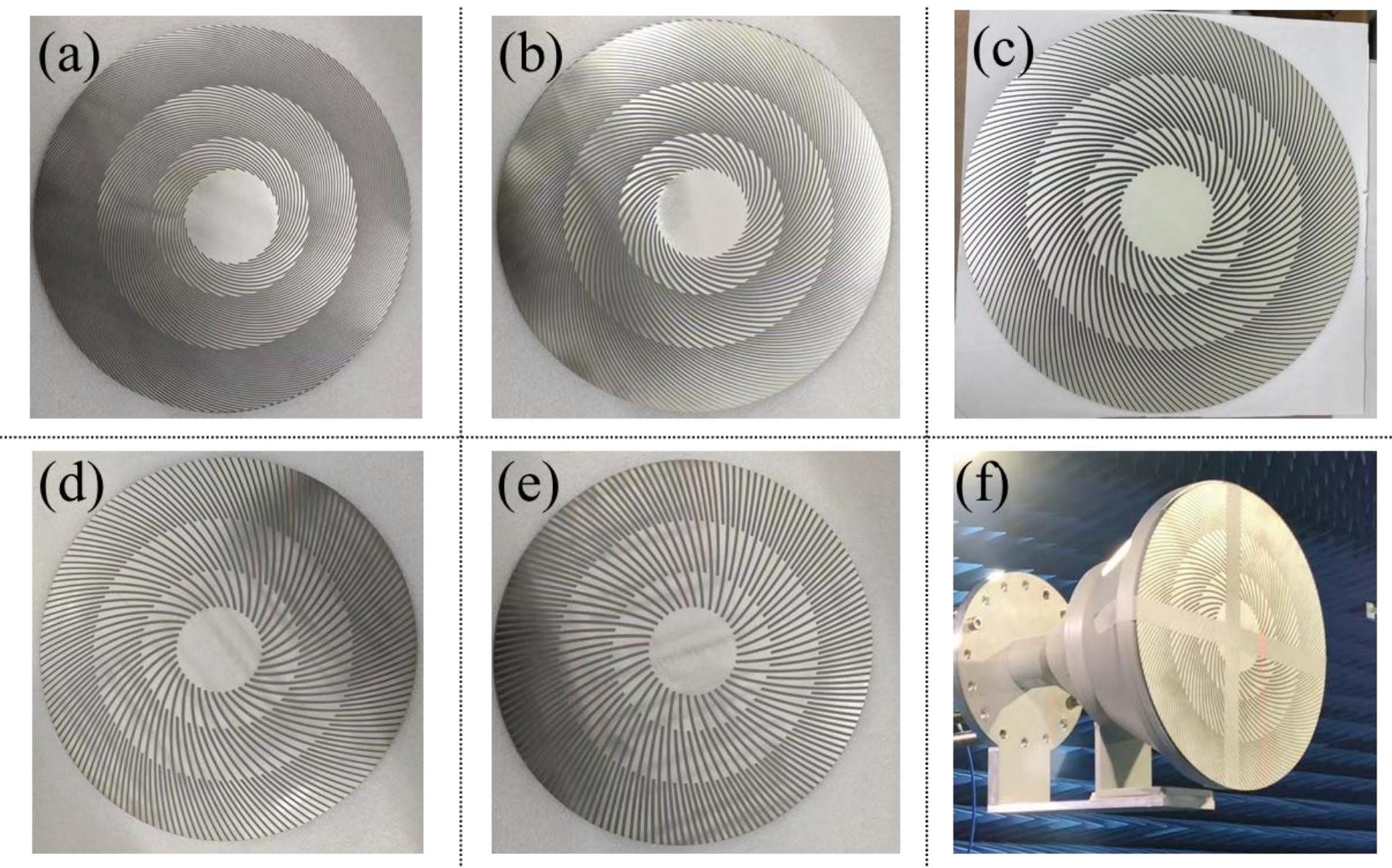

**Fig. S5 Structures of the electromagnetic toroidal helical pulse generator.** The Equiangular spiral grating with (a) $\alpha$ =5π/12, (b) $\alpha$ =π/3, (c) $\alpha$ =π/4, (d) $\alpha$ =π/6, and (e) $\alpha$ =π/12; (f) the equiangular spiral grating fixed to the mouth of the coaxial horn by an adhesive tape. The equiangular spiral grating is composed of 40 sets of slots, each set containing four slots: a long one, a medium-length slot, and two identical short slots.

**Supplementary Note 5: The method for the production of the excitation signal**

The objective of this work is to generate toroidal helical pulses with the specific spatiotemporal distribution and spectral characteristics, represented by the desired spacetime field $y_{idel}(\omega,r)$. This field can be viewed as the outcome of the excitation signal and the spatial response of the electromagnetic toroidal helical pulse generator.

The input signal $x(\omega)$ can be determined after obtaining the spatial response $h(\omega,r)$ of the generator through either simulation or measurement based on the desired field $y(\omega,r)$. To more effectively match the target spacetime field, we choose the signal at a specific location $r_1$ as the reference signal $y_{idel}\left(\omega,r_1^{idel}\right)$, setting $y(\omega,r_1)=$ $y_{idel}\left(\omega,r_1^{idel}\right)$. The input signal that ensures the field at $r_1$ aligns with the theoretical field can be calculated by $x(\omega)=y_{idel}\left(\omega,r_1^{idel}\right)/h(\omega,r_1)$.

In this work, we select the spatial field at $z$ = 40 mm as the target field $y(\omega,r)$, and set $r$ and $r_1^{idel}$ as the locations with maximum measured and theoretical transverse polarization components $E_r$, respectively. After determining the spectral excitation using this method, the time-domain excitation signal is obtained through the inverse Fourier transform.

**Supplementary Note 6: The measurement method for transverse polarized electric fields.**

A planar microwave anechoic chamber was used to measure the transversely polarized spatial electromagnetic fields of the electromagnetic toroidal helical pulse generator, as shown in Fig. S6. The R&S® ZNA vector network analyzer, with a frequency range of 10 MHz to 50 GHz, was connected to both the waveguide probe and electromagnetic toroidal helical pulse generator. The $S_{21}$ measurements were performed to capture the magnitude and phase characteristics of the electromagnetic field at various spatial locations. Given the operational band and mode of the waveguide probe, measurements were conducted across four distinct waveguide bands: 1.7-2.8 GHz, 2.8-3.95 GHz, 3.95-5.85 GHz, and 5.85-8.2 GHz.

The measurements were carried out over the frequency range 1.7-8.2 GHz covered by the four waveguides, with a stepwise frequency increment of 50 MHz. The measurement system allowed the probe to move freely within the measuring space, and it also supported rotation of the antenna for the measurement along its central axis. This procedure enabled direct measurements of the transverse components $E_r$ and $E_\varphi$. The polarization direction of the waveguide probe was adjusted to align with the radial and circumferential directions to measure the $E_r$ and $E_\varphi$ components, respectively.

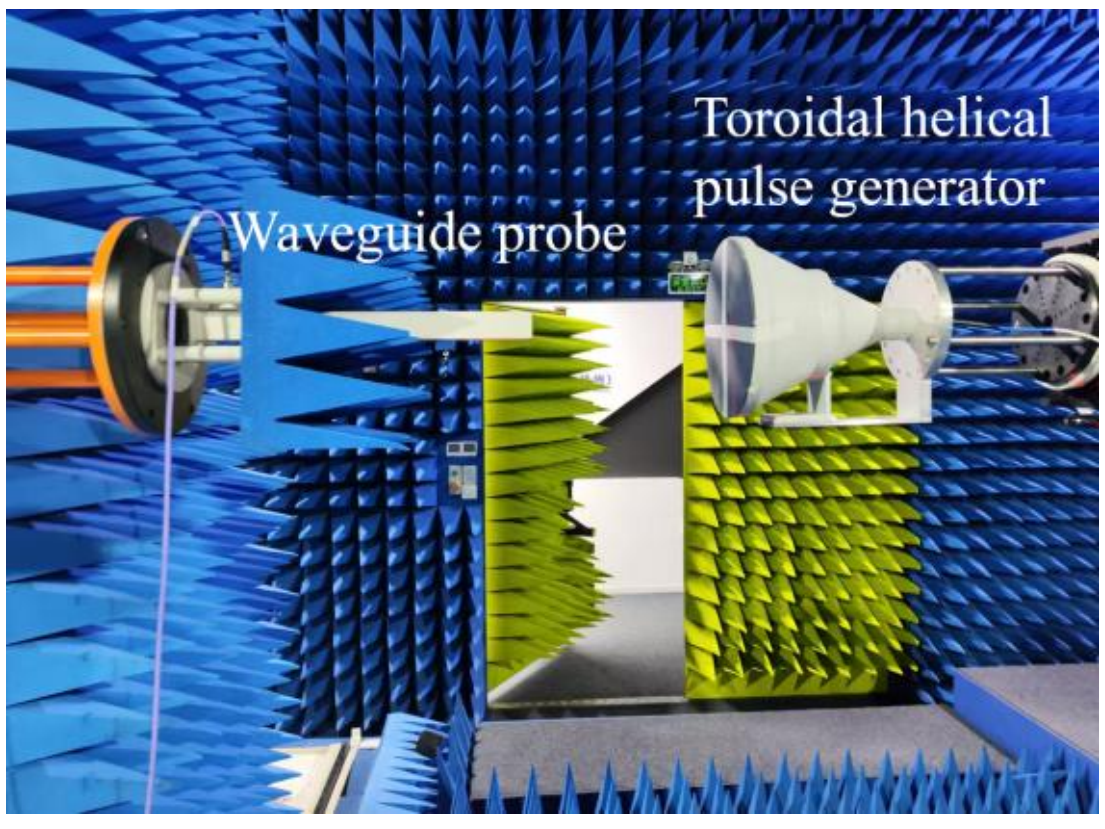

**Fig. S6 The measurement scenario of the electromagnetic toroidal helical pulse generator.** The vector network analyzer was connected to both the waveguide probe and electromagnetic toroidal helical pulse generator. The polarization direction of the waveguide probe was adjusted to align with the radial and circumferential directions to measure the $E_r$ and $E_\varphi$ components, respectively.

The spatial spectra of different transverse polarized electric field components in the planes of $z$ = 20 cm and $z$ = 80 cm of the generated electromagnetic toroidal helical pulse are shown in Fig. S7. The high-frequency and low-frequency components are mainly distributed in the inner and outer regions of the pulse. The $E_\varphi$ and $E_r$ spectral components vanish at the center, these components having similar topologies and shapes, demonstrating the helicity of the vector fields.

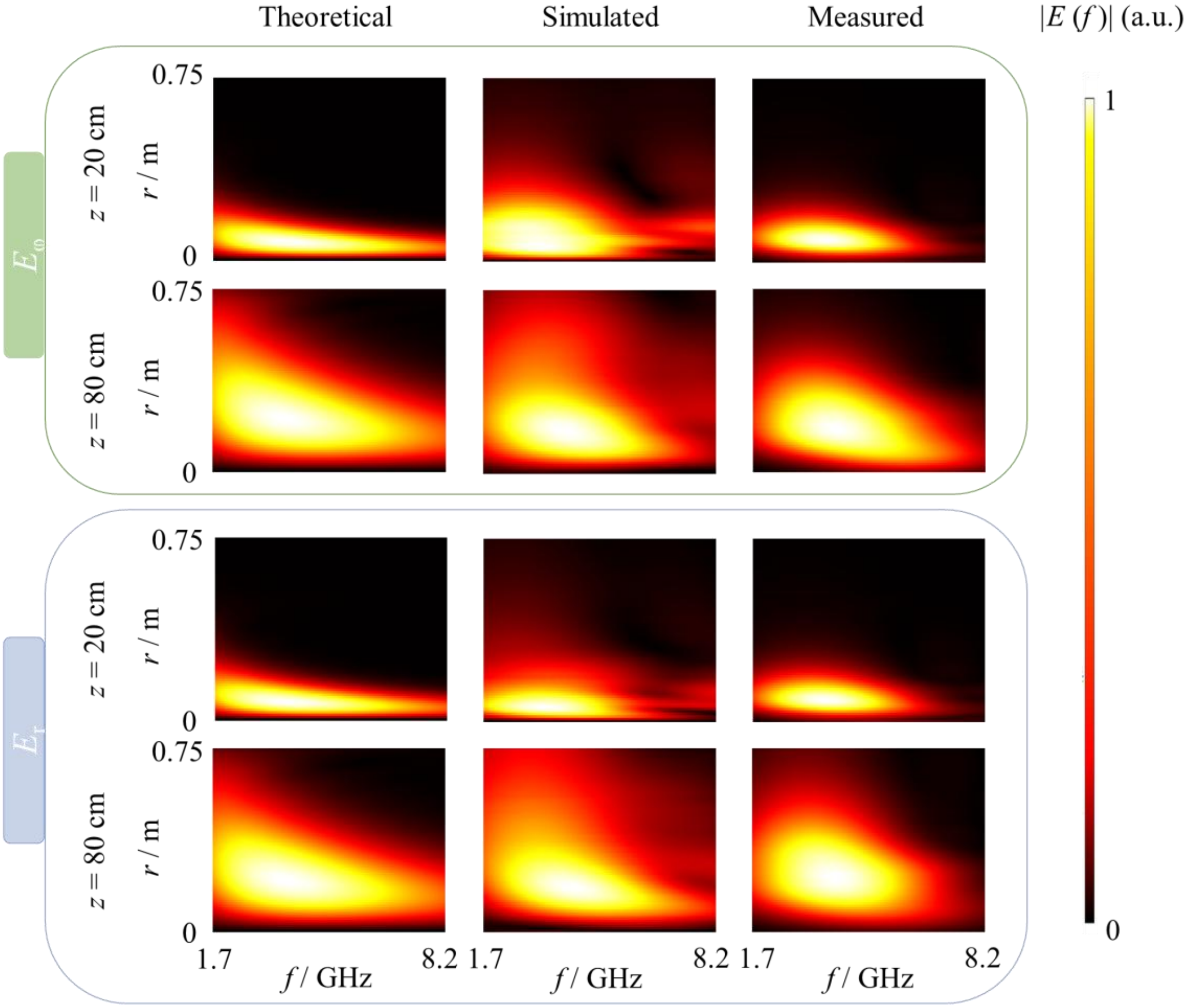

**Fig. S7 The spatial spectra of different transverse polarized electric field components corresponding to the propagating toroidal helical pulses.** The

corresponding parameters are $\alpha = \pi/4$, $q_1 = 0.02$ m, and $q_2 = 20q_1$, and the spatial spectra are plotted in the planes of $z = 20$ cm and $z = 80$ cm. The $E_\varphi$ and $E_r$ spectral components vanish at the center, while the $E_z$ component has its maximum at the center. The spectral distributions of the $E_\varphi$ and $E_r$ components are similar.

As the pulse propagates, the spatial spectra of the electromagnetic toroidal helical pulse show a trend to spread outward. However, the spectral characteristic is preserved, with the high-frequency and low-frequency components staying, predominantly, in the inner and outer regions. The preservation of the spectral distribution in the course of the propagation is a direct consequence of the space-time nonseparability, which is the inherent property of Eqs. (1) – (4).

**Supplementary Note 7: The calculation method for the longitudinal polarized electric fields.**

The longitudinal polarized component can be determined from the measured transverse polarized component, on the basis of the Gauss's law [5], as follows:

$$E_z(r,\varphi,z) = -\int_{z_0}^{z} \left(E_r(r,\varphi,z') + r\frac{\partial E_r(r,\varphi,z')}{\partial r} + \frac{\partial E_\varphi(r,\varphi,z')}{\partial \varphi}\right) dz' . \tag{S28}$$

The spatial spectra of the longitudinal polarized electric field components in the planes of $z = 20$ cm and $z = 80$ cm of the generated electromagnetic toroidal helical pulse are shown in Fig. S8. The theoretical, simulated, and measured $E_z$ spectra of the toroidal helical pulses split in two parts, consistent with the topology displayed in Fig. 1(a), where the strong longitudinally polarized electric fields are located on the inner and outer sides. As the pulse propagates, the longitudinal polarized spatial spectra of the electromagnetic toroidal helical pulse show a trend to spread outward, preserving the splitting spectral characteristic of the longitudinal polarized fields.

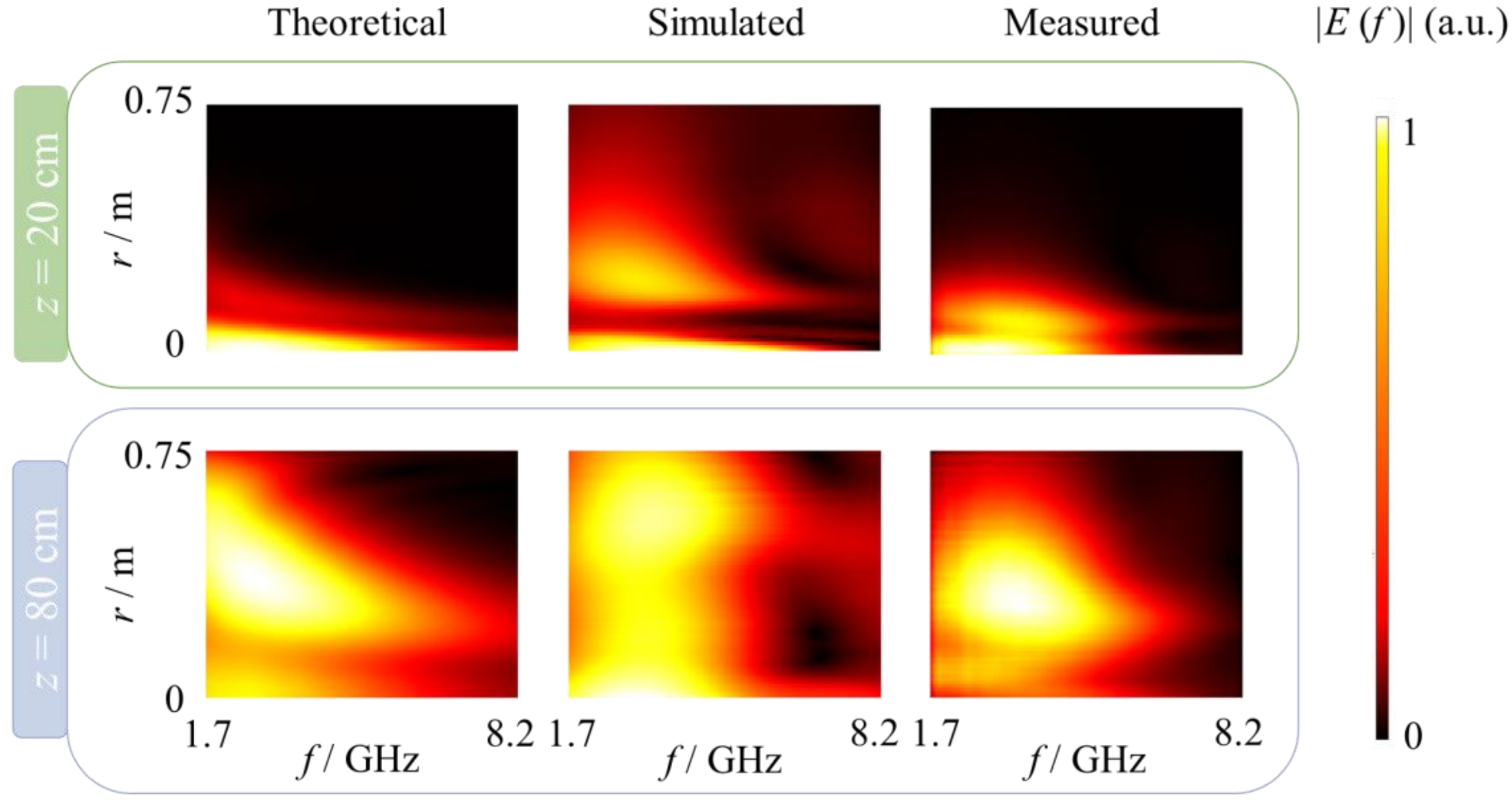

**Fig. S8 The spatial spectra of the longitudinal polarized electric field components corresponding to the propagating toroidal helical pulses.** The corresponding parameters are $\alpha$ =π/4, $q_1$ = 0.02 m, and $q_2$ = 20$q_1$, and the spatial spectra are plotted in the planes of $z$ = 20 cm and $z$ = 80 cm. The $E_z$ spectra of the toroidal helical pulses split in two parts, consistent with the topology displayed in Fig. 1(a), where the strong longitudinally polarized electric fields are located on the inner and outer sides. The splitting spectral characteristic of longitudinal polarized fields is preserved in the course of the pulse propagation.

**Supplementary Note 8: Spatiotemporal electric field of toroidal helical pulses.**

Theoretical and measured three-dimensional spatiotemporal electric fields of toroidal helical pulses are shown in Fig. S9. $E_\varphi$ and $E_r$ components have obvious two or three lobes, indicating a single or 1.5 cycle feature. Fields $E_\varphi$ and $E_r$ vanish at the center, the $E_z$ component has its maximum at the same place, indicating the longitudinal polarization in the central area. The $E_\varphi$ and $E_r$ components feature similar topologies and shapes, demonstrating the helicity of the vector fields. The simulated spatiotemporal fields may exhibit additional cycles in comparison to the theoretical ones. For instance, two lobes are observed in Fig. S9(c1), while three lobes appear in Fig. S9(c2). This variation is attributed to the limited bandwidth, which can be seen in detail from the spectrum in Fig. S7.

The spatiotemporal electric fields in longitudinal cross-sections of the toroidal helical pulses generated during the propagation are shown in Fig. S10. The $E_\varphi$ and $E_r$ components demonstrate that, in the course of the propagation, the toroidal helical pulses gradually evolve from 1.5 cycles to the single cycle. The simulated and measured results demonstrate the evolution similar to the theoretical results. Similar to Fig. S9, the $E_\varphi$ and $E_r$ components vanish at the center in all planes; outside of the center, these components exhibit similar topologies and shapes, demonstrating the helicity of the vector fields. Furthermore, unlike the $E_\varphi$ and $E_r$ components, the theoretical, simulated, and measured $E_z$ components of the toroidal helical pulses show a trend to split into two fragments in the radial direction, which is consistent with the vector field distribution observed in Fig. 3. This fact indicates that the strong longitudinally polarized electric fields are located on the inner and outer sides of the toroid that are encircled by the vector field. The transverse fields ($E_\varphi$ and $E_r$) and the longitudinal one ($E_z$) together form the toroidal helical topology, which can also be directly observed in the vector fields in the transverse planes, as shown in Fig. 3.

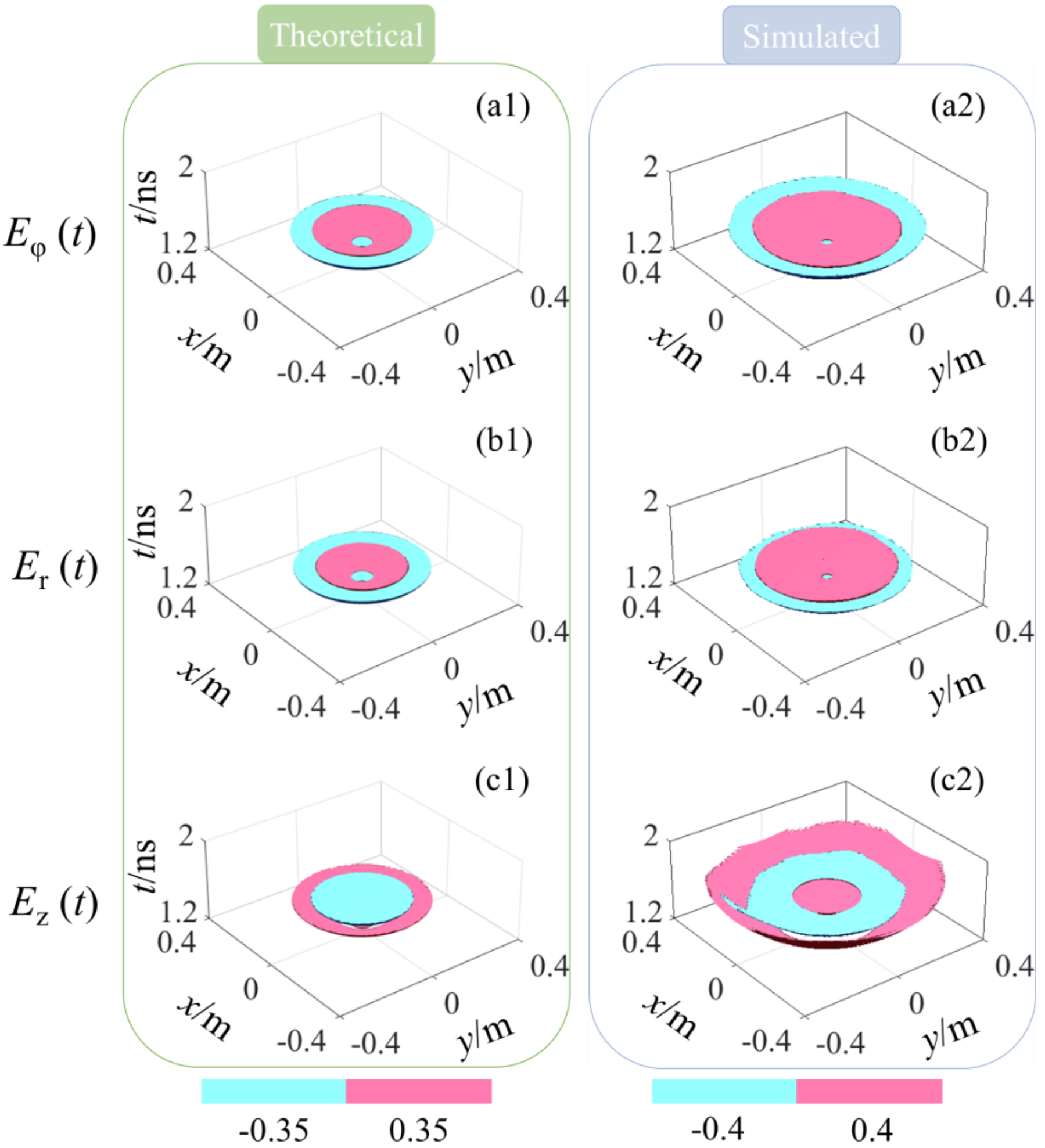

**Fig. S9 Three-dimensional spatiotemporal electric field in a generated electromagnetic toroidal helical pulse with the parameter of $\alpha$ =π/4.** Panels (a1-a2), (b1-b2), and (c1-c2) display the $E_\varphi$, $E_r$, and $E_z$ components, respectively. (a1-c1) and (a2-c2) are theoretical and simulated results. The corresponding parameters are $q_1$=0.02 m and $q_2$=20$q_1$, and the spatiotemporal electric fields are displayed in the plane of $z$ = 40cm. Fields $E_\varphi$ and $E_r$ vanish in the center, where the $E_z$ component attains its

maximum, indicating the longitudinal polarization at this spot. The $E_\varphi$ and $E_r$ components feature similar topologies and distributions, demonstrating the helicity of the vector fields.

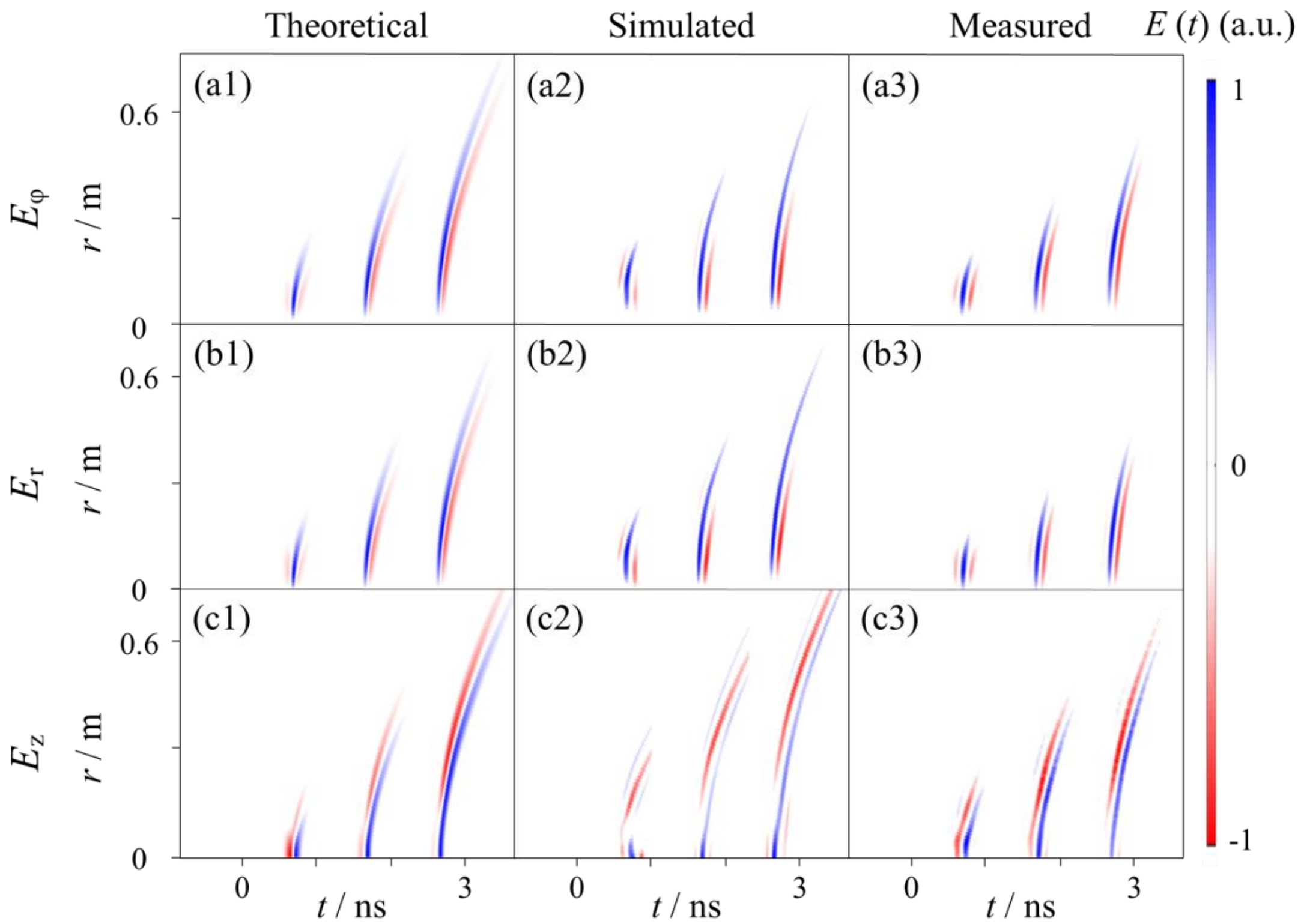

**Fig. S10 The spatiotemporal electric field of the generated toroidal helical pulses in the (*x*, *t*) plane.** The corresponding parameters are $\alpha = \pi/4$, $q_1 = 0.02$ m, and $q_2 = 20q_1$. Panels (a1-a3), (b1-b3), and (c1-c3) display the $E_\varphi$, $E_r$, and $E_z$ components, respectively, while panels (a1-c1), (a2-c2), and (a3-c3) represent the theoretical, simulated, and measured results. Each subfigure plots the spatiotemporal electric fields in the planes of $z = 20$ cm, $z = 50$ cm, and $z = 80$ cm, to display the propagating process. The $E_\varphi$ and $E_r$ components demonstrate that, in the course of the propagation, the toroidal helical pulses gradually evolve from the 1.5 cycles towards the single cycle.

**Supplementary Note 9: The calculation method for the skyrmion number ($N_{sk}$) and direction sphere.**

The topological properties of the 2D skyrmion model are characterized by integral $N_{sk}$ of the topological charge density over space [6], which is a topological invariant defined as:

$$N_{Sk} = \int n_{Sk}(\boldsymbol{r}) d^2 r \,, \tag{S29}$$

$$n_{Sk}(\boldsymbol{r}) = \frac{1}{4\pi} \mathbf{m}(\mathbf{r}) \cdot \left[ \frac{\partial \mathbf{m}(\mathbf{r})}{\partial x} \times \frac{\partial \mathbf{m}(\mathbf{r})}{\partial y} \right], \tag{S30}$$

where $\mathbf{m}(\boldsymbol{r}) = \frac{\mathbf{E}(\boldsymbol{r})}{|\mathbf{E}(\boldsymbol{r})|}$ is the unit vector that constructs the skyrmion pattern, and $\mathbf{E}(\boldsymbol{r})$ is the electric field vector at $\boldsymbol{r} = (x, y)$. The spatial domain fixed for the calculations is determined by the size of the skyrmion.

Using the azimuthal angle $\alpha$ and polar angle $\beta$ to represent the vector basis in spherical coordinates, and expressing the position vector in polar coordinates as $\boldsymbol{r} = r(\cos\varphi, \sin\varphi)$. Based on this mapping, the vector field can be written as $\boldsymbol{m} = (\cos\alpha(\varphi)\sin\beta(r), \sin\alpha(\varphi)\sin\beta(r), \cos\beta(r))$, then Eq. (S29) can be rewritten as

$$\begin{aligned} n_{Sk}(\boldsymbol{r}) &= \frac{1}{4\pi} \int_0^{r_\sigma} dr \int_0^{2\pi} d\varphi \frac{d\beta(r)}{dr} \frac{d\alpha(\varphi)}{d\varphi} \sin\beta(r) \\ &= \frac{1}{4\pi} \left[\cos\beta(r)\right]_{r=0}^{r=r_\sigma} \left[\cos\alpha(\varphi)\right]_{\varphi=0}^{\varphi=2\pi} = pm \end{aligned}, \tag{S31}$$

The integer $p = \left[\cos\beta(r)\right]_{r=0}^{r=r_\sigma} / 2$ defines the polarity. When $p = 1$ ($p = -1$), the vector at the center $r = 0$ points downward (upward), while at the boundary $r = r_\sigma$ it points upward (downward). The integer $\left[\alpha(\varphi)\right]_{\varphi=0}^{\varphi=2\pi} / 2\pi$ defines the vorticity, which controls the distribution of the transverse field components.

To distinguish different helicities of the vortex, an initial phase $\gamma$ is introduced into $m$ : $\alpha(\varphi)=m\varphi+\gamma$ , Under the parameter condition $p=1$ the skyrmion structures as functions of $m$ and $\gamma$ are shown in Fig. S11. Panels (a) and (b) are referred to as Néel-type and Bloch-type skyrmions, respectively.

When $m=1$ the configuration is a skyrmion; when $m=-1$, it is an antiskyrmion. From the perspective of the transverse component angles, the skyrmion structure may be vortex-like, radial, or intermediate between the two, whereas the antiskyrmion structure exhibits a quadrupolar pattern. The parameter $\gamma$ determines the rotation angle of the vector arrows. Taking Fig. S11(a) and (b) as examples, a skyrmion with $\gamma=\pi/2$ can be obtained by rotating all vectors of theskyrmion counterclockwise by $\pi/2$ which also reflects the distinction between vortex-like and radial configurations. For panels (c) and (d), the skyrmion with $\gamma=\pi/2$ can likewise be obtained by rotating the $\gamma=0$ configuration; however, due to the intrinsic quadrupolar structure, each vector remains consistent with the quadrupolar pattern after rotation.

The helicity of the skyrmion is defined as

$$\gamma=\theta-Q\phi\,, \tag{S32}$$

where $\theta$ and $\phi$ denote the polar and azimuthal angles of the vector direction, respectively, and $Q=\pm1$ is the topological charge.

To obtain the actual helicity from experimental and simulation data, we first determine the skyrmion center and then calculate the average helicity of all vectors within the selected region:

$$\gamma_{\text{avg}}=\frac{1}{N}\sum_{i=1}^{N}\theta_i-\phi_i\ , \tag{S33}$$

where $N$ is the total number of vectors within the selected region.

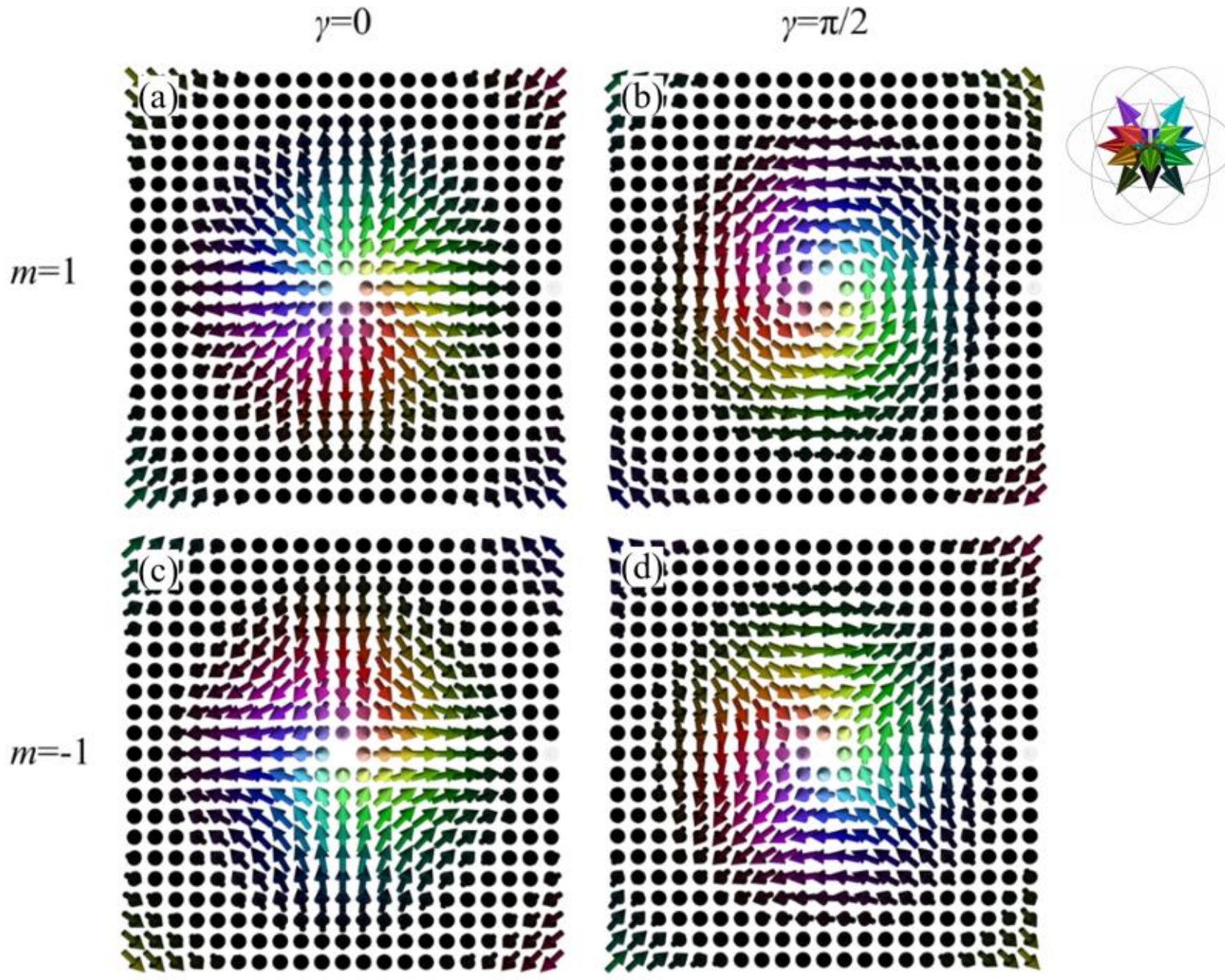

**Fig. S11 Skyrmion field structures under different parameter conditions.** The arrow colors correspond to the polar and azimuthal angles in the spherical coordinate system.(a) Néel type; (b) Bloch type; (c) $m=-1$, $\gamma=0$; (d) $m=-1$, $\gamma=\pi/2$

The helicity estimation results for all skyrmion structures are plotted in Fig. 3 (left vertical axis). The results show that, except for a few states that exhibit predictable deviations from the ideal target, the remaining measurements and simulations are in very good agreement with the theoretical helicity values.

The vector electric fields in the transverse plane of the toroidal helical pulses with $\alpha$ = 5π/12 and π/12 are shown in Fig. S12. The longitudinal component of the field gradually decreases from the center outward, reverses direction after passing through zero, and then gradually increases again. Meanwhile, the transverse component maintains an angle of approximately $\alpha$ relative to the radial direction. The position of the maximum value of the longitudinal component coincides with the zero point of the transverse component. These features indicate that the vector field of the electromagnetic toroidal helical pulse in the transverse planes exhibits a hybrid skyrmion texture, combining Neel and Bloch skyrmions.

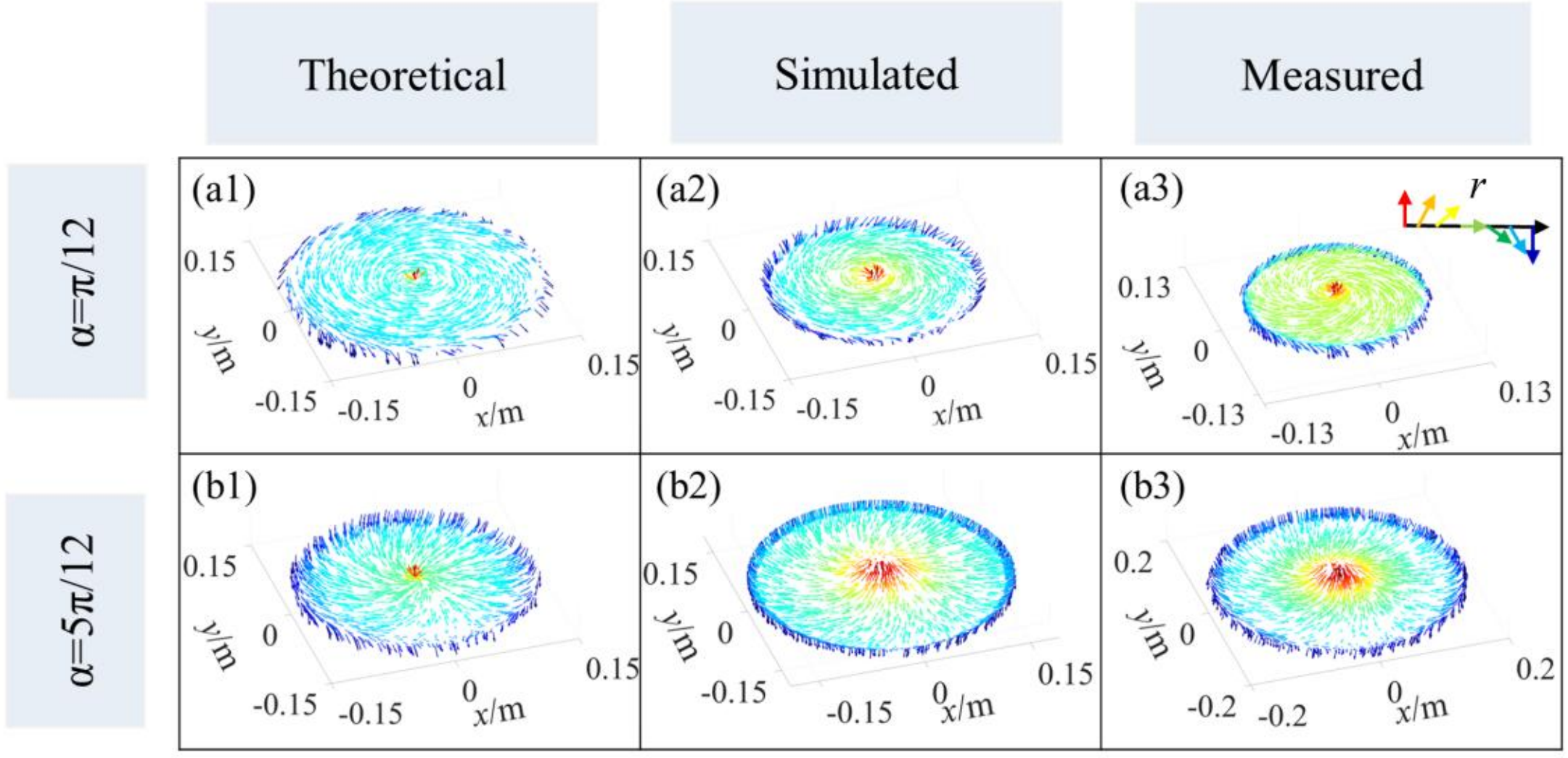

**Fig. S12 The vector electric field of electromagnetic toroidal helical pulse in the transverse planes.** The vector fields in (a1, b2), (a2, b2), and (a3, b3) are shown in the planes of $z$ = 40 cm at $t$ = 1.32 ns, $t$ = 1.39 ns, and $t$ = 1.38 ns, respectively. The vector field of the electromagnetic toroidal helical pulse in the transverse planes exhibits a hybrid skyrmion texture, combining Neel and Bloch skyrmions.

To examine the propagation stability of the skyrmion texture, the transverse skyrmion field distributions of the theoretical toroidal helical pulse were calculated at propagation distances of $z$ = 0.4 m, 0.8 m, and 1.2 m, as shown in Fig. S13. The results indicate that the skyrmion topology remains preserved during propagation. As the pulse propagates in free space, the overall field distribution undergoes diffraction-induced expansion, while the internal polarization mapping associated with the skyrmion structure remains unchanged. This behavior confirms that the skyrmion texture evolves in a self-similar manner, exhibiting only uniform scaling without distortion or topological transformation.

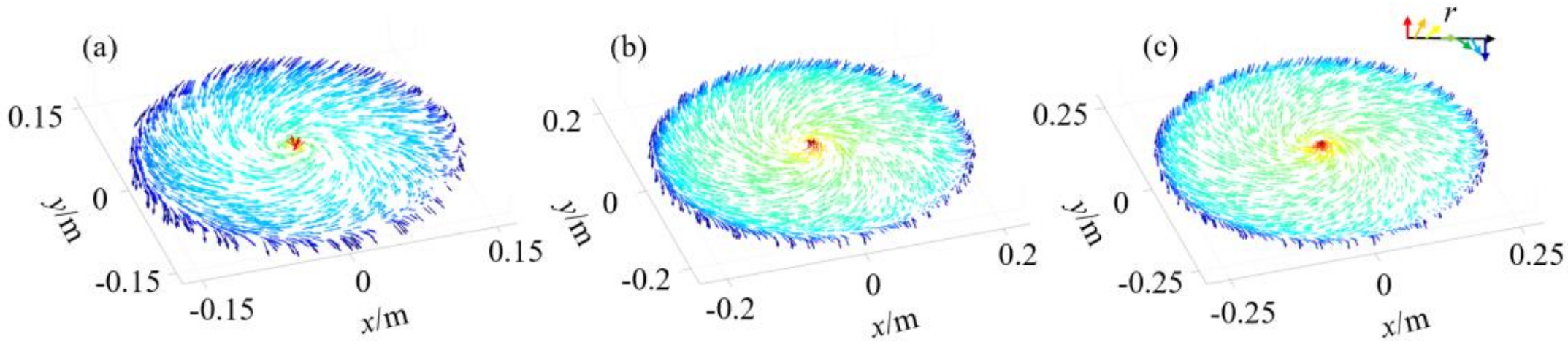

**Fig. S13 Transverse skyrmion field distributions of the theoretical toroidal helical pulse at different propagation distances**. (a) $z$ = 0.4 m, (b) $z$ = 0.8 m, and (c) $z$ = 1.2 m. The results show that the skyrmion texture preserves its topological structure during propagation, while the overall field distribution undergoes diffraction-induced expansion.

The vector characteristics of skyrmions can also be visually shown by mapping the directions of vectors onto a sphere of directions (with radius 1), using Cartesian sampling of a skyrmion's cell [6]. The position on the direction sphere of a vector is determined by its own direction. After normalizing the length of the vector to 1, the vector is mapped from the origin to the point on the sphere's surface that corresponds to its direction. When the skyrmion number is an integer, the vector field in the target region fully covers the surface of the direction sphere, as shown in Fig. S14.

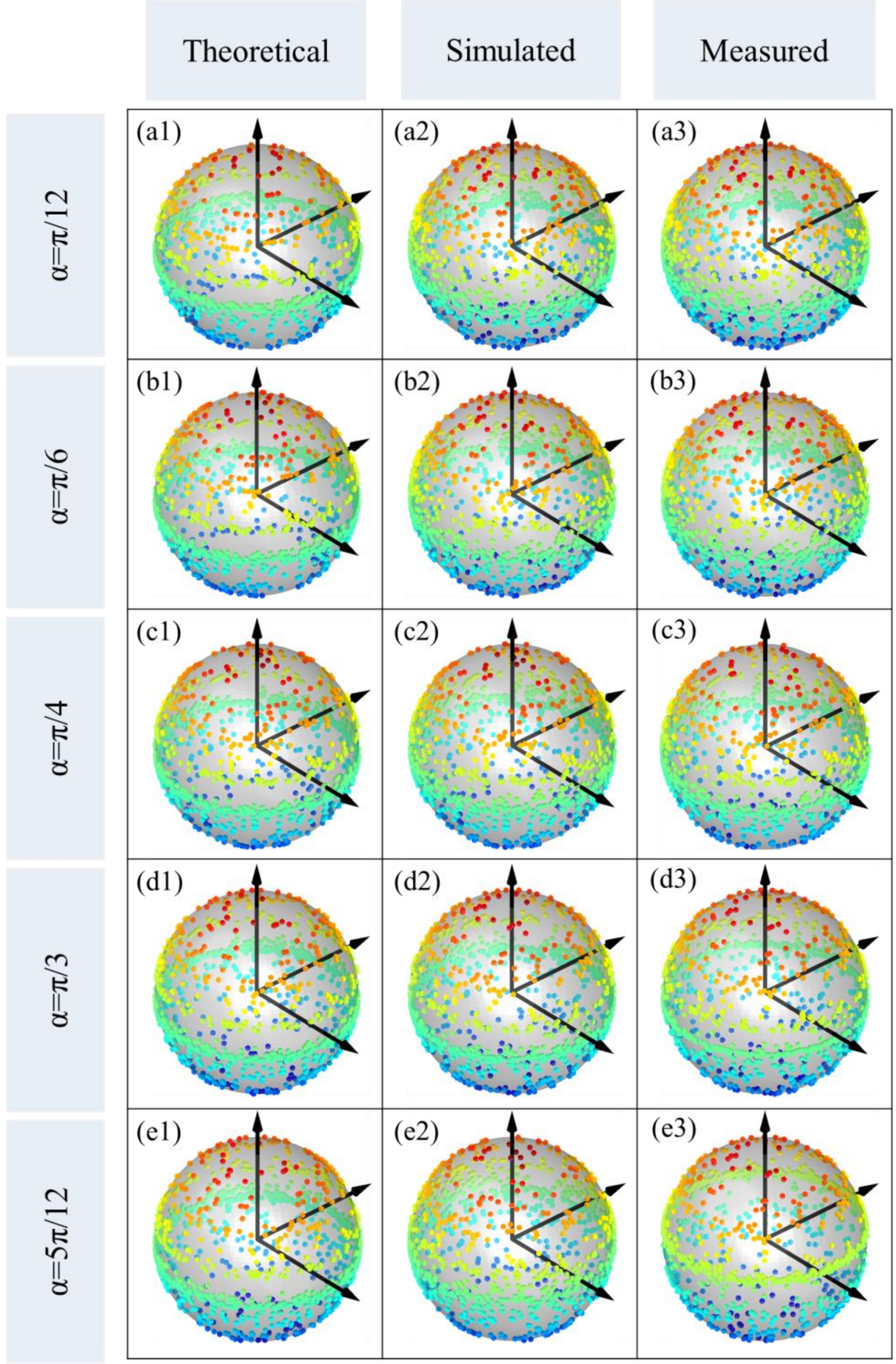

**Fig. S14 The direction sphere of the vector field in the transverse planes of toroidal helical pulses with (a) $\alpha$ =π/12, (b) $\alpha$ =π/6, (c) $\alpha$ =π/4, (d)$\alpha$ =π/3, and (e)$\alpha$ =5π/12.** The direction of the vector field is distributed across the entire sphere, confirming the skyrmion texture.

**Supplementary Note 10: The calculation method for the space-time nonseparability.**

The space-time nonseparability of pulses manifests itself as an isodiffracting characteristic, meaning that, in all cross-sections perpendicular to the propagation direction, the spatial intensity distribution of each frequency component diffuses along the pulse's trajectory in the same manner, as shown in Figs. S15(a1-a2). The maximum-amplitude trajectories of different frequency components do not intersect. Based on the method from Ref. [7], some concepts can be borrowed from quantum mechanics to quantitatively characterize the space-time nonseparability of the electromagnetic toroidal helical pulse.

(1) Concurrence (Con): Con is a continuous measure of the nonseparability between entangled states, defined as $C=\sqrt{2\left[1-\mathrm{Tr}\left(\rho_A^2\right)\right]}$, where $\rho_A$ is the reduced density matrix.

(2) Entanglement of formation (Eof): Eof is another commonly used measure of quantum entanglement. It is computed from the reduced density matrix as $E=-\mathrm{Tr}\left(\rho_A \log_2\left(\rho_A\right)\right)$.

Both Con and Eof serve as measures of the nonseparability. The normalized Con and Eof values range from 0 to 1: 0 indicates complete separability, and 1 indicates the maximum of the nonseparability. For toroidal helical pulses, the entangled states used to calculate Con and Eof correspond to the spectral and spatial states. The spectral states $|\lambda_i\rangle (i=1,2,.....n)$ define monochromatic states with peak intensity at points $\mathbf{r}_{\lambda i}$ corresponding to wavelength $\lambda_i$. The spatial states $|\eta_i\rangle$ represent polychromatic states at $\boldsymbol{\eta}_i = \mathbf{r}/r_{max}$, where $r_{max}$ is the radial position of the maximum total field intensity.

Based on this definition, the computed values of Con and Eof of the toroidal helical pulses show fluctuations near $z$ = 0 and remain above 0.85 at $z$ = 0.4 m, as shown in Figs. S15(b1-b2), indicating a strong isodiffracting characteristic and spatiotemporal nonseparability.

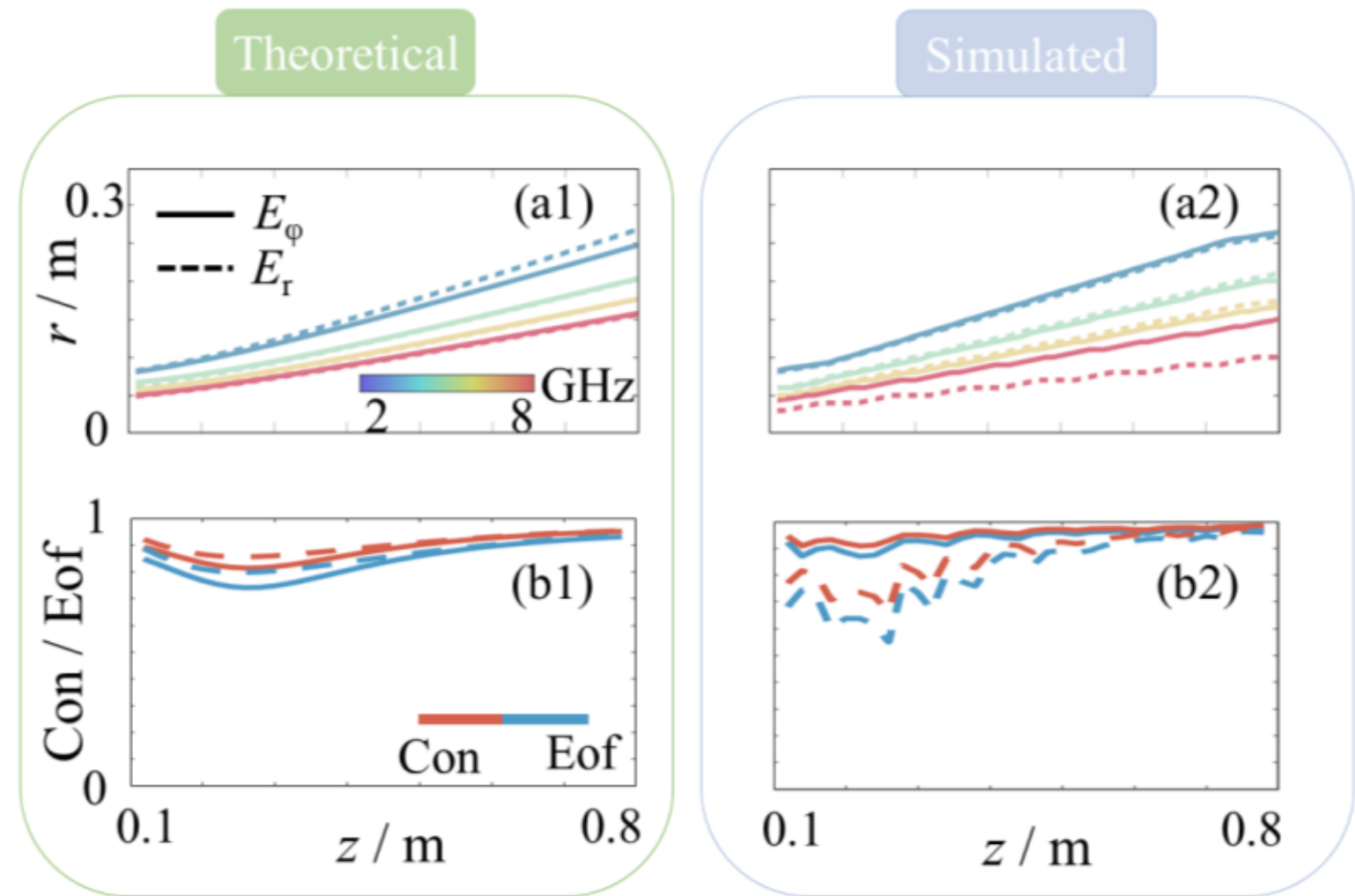

**Fig. S15 Theoretical and simulated space-time nonseparability of a electromagnetic toroidal helical pulse with a typical parameter of $\alpha$ = $\pi$/4.** The corresponding parameters are $q_1$=0.02 m and $q_2$=20$q_1$. The theoretical and simulated trajectories of the maximum spectrum of different components of the generated toroidal pulse during the propagation are shown in (a1-a2). The trajectories of the $E_\varphi$ and $E_r$ components exhibit similar behavior and remain non-intersecting, demonstrating isodiffracting characteristics, which are directly evaluated in panel (b1-b2) in terms of the concurrence (Con) and entanglement of formation (Eof). At $z$ > 0.4 m, the Con and Eof values of the generated toroidal helical pulses remain above 0.85.